\documentclass[twocolumn]{aastex631}

\usepackage{amsmath}
\usepackage{textgreek}
\usepackage{xcolor}
\graphicspath{{figs/}}

\DeclareMathSymbol{\leq}{\mathrel}{symbols}{20}
   
\DeclareMathSymbol{\geq}{\mathrel}{symbols}{21}

\usepackage{booktabs}
\usepackage{xspace}
\newcommand{\code}[1]{\textsc{#1}}
\newcommand{\ascent}{\code{Ascent}\xspace}
\newcommand{\athenapk}{\code{AthenaPK}\xspace}
\newcommand{\athenapp}{\code{Athena++}\xspace}
\newcommand{\parthenon}{\code{Parthenon}\xspace}
\newcommand{\kokkos}{\code{Kokkos}\xspace}

\newcommand{\simid}[1]{\texttt{#1}}
\newcommand{\fid}{\simid{Fiducial}\xspace}
\newcommand{\weakB}{\simid{WeakMag}\xspace}
\newcommand{\weakC}{\simid{WeakCool}\xspace}
\newcommand{\hydro}{\simid{Hydro}\xspace}

\begin{document}

\title{The XMAGNET exascale MHD simulations of SMBH feedback in galaxy groups and
  clusters: Overview and preliminary cluster results}

\author[0000-0003-3555-9886]{Philipp Grete}
\affiliation{University of Hamburg, Hamburger Sternwarte, Gojenbergsweg 112, 21029 Hamburg, Germany}
\author[0000-0002-2786-0348]{Brian W.\ O'Shea}
\affiliation{Department of Computational Mathematics, Science, \& Engineering, Michigan State University, 428 S. Shaw Lane, East Lansing, MI 48824}
\affiliation{Department of Physics \& Astronomy, 567 Wilson Road, Michigan State University, East Lansing, MI 48824}
\affiliation{Facility for Rare Isotope Beams, Michigan State University, 640 S. Shaw Lane, East Lansing, MI 48824}
\affiliation{Institute for Cyber-Enabled Research, 567 Wilson Road, Michigan State University, East Lansing, MI 48824}

\author[0000-0002-6837-8195]{Forrest W. Glines}
\affiliation{Theoretical Division, TA-3 Bldg. 123, Los Alamos National Laboratory, Los Alamos, NM 87545}

\author[0000-0003-1255-6375]{Deovrat Prasad}
\affiliation{School of Physics and Astronomy, Cardiff University, 5 The Parade, Cardiff CF24 3AA, UK}

\author[0000-0003-3175-2291]{Benjamin D. Wibking}
\affiliation{Department of Physics \& Astronomy, 567 Wilson Road, Michigan State University, East Lansing, MI 48824}

\author[0009-0006-2593-1583]{Martin Fournier}
\affiliation{University of Hamburg, Hamburger Sternwarte, Gojenbergsweg 112, 21029 Hamburg, Germany}

\author[0000-0002-3369-7735]{Marcus Br\"uggen}
\affiliation{University of Hamburg, Hamburger Sternwarte, Gojenbergsweg 112, 21029 Hamburg, Germany}

\author[0000-0002-3514-0383]{Mark Voit}
\affiliation{Department of Physics \& Astronomy, 567 Wilson Road, Michigan State University, East Lansing, MI 48824}



\begin{abstract}

We present initial results from extremely well-resolved 3D magnetohydrodynamical simulations of idealized galaxy clusters, conducted using the AthenaPK code on the Frontier exascale supercomputer.
These simulations explore the self-regulation of galaxy groups and cool-core clusters by cold gas-triggered active galactic nucleus (AGN) feedback incorporating magnetized kinetic jets.
Our simulation campaign includes simulations of galaxy groups and clusters with a range of masses and intragroup and intracluster medium properties. In this paper we present results that focus on a Perseus-like cluster.
We find that the simulated clusters are self-regulating, with the cluster cores staying at a roughly constant thermodynamic state and AGN jet power staying at physically reasonable values ($\simeq 10^{44}-10^{45}$~erg/s) for billions of years without a discernible duty cycle. These simulations also produce significant amounts of cold gas, with calculations having strong magnetic fields generally both promoting cold gas formation and allowing cold gas out to much larger clustercentric radii ($\simeq 100$~kpc) than simulations with weak or no fields ($\simeq 10$~kpc), and also having more filamentary cold gas morphology. We find that AGN feedback significantly increases the strength of magnetic fields at the center of the cluster. We also find that the magnetized turbulence generated by the AGN results in turbulence where the velocity power spectra are tied to AGN activity whereas the magnetic energy spectra are much less impacted after reaching a stationary state.


\end{abstract}

\keywords{Galaxy clusters (584), Galaxy Jets (601), Intracluster
  medium (858), Magnetic fields (994)}


\section{Introduction}\label{sec:intro}


The intracluster medium (ICM) -- the hot, diffuse, magnetized plasma that comprises the bulk of the baryons in galaxy clusters -- radiates copiously in X-ray radiation, with X-ray luminosities of massive clusters typically being in the range of $\sim 10^{43.5} - 10^{45.5}$~erg/s \citep{Fabian_1994}.  ``Cool core'' galaxy clusters, having dense and (relatively) cool cores, are particularly X-ray luminous, with central cooling times of tens to hundreds of millions of years -- far shorter than the age of the universe and the lifetime of these structures \citep{Hudson_2010, Cavagnolo2008ApJ}.

In cool-core clusters, local thermal stability of the intracluster medium becomes a critical physical phenomenon that governs their evolution \citep{Field1965ApJ}.  Cold gas is expected to precipitate out of the thermally unstable ICM \citep{Voit_2008,McCourt2012MNRAS,Sharma2012} and to be available to star formation and accretion onto the central SMBH \citep{soker2001,Oosterloo_2024,Guo_2024}.
An apparent consequence is that the rapidly cooling gas would result in many hundreds to thousands of solar masses per year of cold, dense gas being available for star formation. This would result in brightest cluster galaxies having a very high star formation rate ($\sim$ several $100s$ M$_\odot$ yr$^{-1}$), and thus
it was predicted that the brightest central galaxies in clusters should see similarly high star formation rates. The observed  star formation rate in these galaxies is lower, however -- typically only a handful of solar masses per year \citep{Fabian1984Nature,Edge2001MNRAS,Peterson2003ApJ,ODea_2008}.

This observation implies that there is a heating source in the center of every cool-core galaxy cluster that offsets this cooling, and does so in a tightly-coupled way that inhibits significant amounts of gas from forming \citep{Pizzolato2005ApJ}.  Although several mechanisms have been proposed, including thermal conduction \citep{Voigt2004, Ruszkowski2011}, supernova feedback \citep{voit2001Natur}, sound waves \citep{2004ApJ...611..158R}, 
infalling galaxies and sub-halos \citep{Dekel2008}, and cluster mergers, 
the only mechanism that robustly fits the necessary criteria of (1) appropriate level of energy production, (2) location that would tend to deposit heat in the areas of greatest X-ray emission/cooling, and (3) ability to couple to the central cluster plasma on short time scales, is AGN feedback \citep{Churazov2001,Birzan2004ApJ,Rafferty2006ApJ,Rafferty2008ApJ,Cavagnolo2008ApJ,Cavagnolo2010ApJ,Sun2009ApJ}. 
This idea has been supported by observational evidence from radio surveys showing that radio-loud AGN at the centers of massive cool-core galaxies are always active and, separately, analysis of cavities created by AGN jets interacting with the ICM demonstrates that the work done by those ``bubbles'' over time is roughly what is needed to offset cooling in clusters \citep[e.g.,][]{Birzan2004ApJ,Panagoulia2014MNRAS,Prunier_2024}. 

These observations are supported by theory and simulations that have shown AGN feedback is crucial to the self-regulation of clusters
\citep[e.g.,][]{Gaspari2012ApJ,Li2015ApJ,Prasad2015ApJ,Yang2016ApJ,Meece2017ApJ,Mila2021,Nobels2022MNRAS,Ehlert2023MNRAS}.  
Idealized simulations of AGN feedback in galaxy clusters have demonstrated that cold gas-triggered AGN hosted by the central cluster galaxy are effective mechanisms for offsetting radiative losses in the intracluster medium, and (depending on the exact prescription for AGN triggering and feedback) result in clusters with thermodynamic states that are comparable to observed cool-core clusters \citep[e.g.,][]{Li_2014,Qiu_2019,Beckmann_2019,Wang_2021,Ehlert2023MNRAS,Fournier2024A&A}.
Furthermore, theory has shown that AGN feedback is
important in environments less extreme than galaxy clusters. Cosmological simulations 
 of galaxy formation show that AGN feedback is necessary to correctly reproduce the low star formation rates in massive galaxies \citep[e.g.,][]{Naab2017ARA&A,2018MNRAS.479.4056W,DonahueVoit2022PhR,2022MNRAS.512.1052P}
as well as the budget of cold-phase gas \citep[e.g.,][]{Olivares2019A&A}. It is challenging to make detailed inferences from these simulations, however, because  implementations of AGN feedback operate at the resolution limit of these calculations (i.e., on $\sim$ kpc scales) and thus cannot accurately capture the interaction between the jet and the cooling, multi-phase gas. In addition, these simulations do not reproduce the morphology of the radio jets that dominate the radio sky \citep[e.g.,][]{Shimwell2022A&A}.

It is clear that there is need for a detailed examination of the impact of AGN feedback in self-regulating cool-core clusters.  
Multi-wavelength observations of these systems have recently shed light on the multiphase nature of the intracluster medium, showing significant complexity at relatively small spatial scales \citep[e.g.,][]{Salome2006A&A,McNamara2014ApJ,Tremblay2018ApJ,Olivares2019A&A,Vantyghem2021ApJ,Gingras2024ApJ}.
Filament-like structures of magnetically-supported warm ionized and cold molecular gas are expected to populate the inner tens of kpc of potentially all cool--core clusters \citep{Russell2014ApJ,Olivares2019A&A}, with cold molecular gas also
sometimes observed in much smaller, disk-like structures \citep[see, e.g.,][]{Hamer2014}. The measured velocity structure functions and X-ray surface brightness fluctuations also imply that the ICM is turbulent, although this turbulence is subsonic \citep{Li2020ApJ,Ganguly2023,deVries2023MNRAS}.

While subsonic (and thus subdominant in terms of the system's overall pressure support), this turbulence may be quite important for promoting thermal instability and energy transfer.  Highly idealized simulations using Cartesian, turbulent boxes have shown that the formation of cold structures and the coupling between gas phases is tightly coupled to turbulence driving, magnetic fields, and potentially other physical processes \citep{McCourt2012MNRAS,Gaspari2013MNRAS,Ji2018MNRAS,Mohapatra2022MNRAS,Mohapatra2023MNRAS,Wibking2024arXiv241003886W}.
 While these setups are useful for parametric studies and provide a way to probe the effects of turbulence in hot plasmas, they rely on simplified models of turbulence driving, missing the anisotropic and intermittent nature of AGN feedback as well as the impact of the overall cluster environment.
 
 In this paper we present the first results from the XMAGNET (``eXascale simulations of Magnetized AGN feedback focusing on Energetics and Turbulence'') project\footnote{See \url{https://xmagnet-simulations.github.io} for further material including videos.} --  
 a suite of very high resolution 3D magnetohydrodynamical simulations of idealized galaxy groups and clusters that were run on Frontier\footnote{Frontier is operated by the Oak Ridge National Laboratory's Leadership Computing Facility on behalf of the Department of Energy. This work is supported by the DOE INCITE program under allocation AST-146 (2023-2024).}, the first exascale supercomputer available to academic researchers~\citep{Atchley2023}.  These simulations, run using the AthenaPK code, include gravity, radiative cooling, and a model of cold gas-triggered AGN feedback that converts accreted gas into bipolar magnetized kinetic jets and thermal energy.
 The goal of this simulation suite is to understand, using extremely large dynamic range and physics-rich calculations with very well-resolved core regions, the accretion of gas onto supermassive black holes, the impact of the resulting magnetized AGN jets on the circumgalactic and intracluster plasma. By resolving the inner regions of these idealised systems with extremely large uniform-resolution grids, we attempt to bridge the gap between realistic but relatively poorly resolved cosmological simulations of groups and galaxies \citep[e.g.,][]{Pellissier2023MNRAS,Schaye2023MNRAS,Nelson2024A&A} and high-resolution but also highly idealized turbulent boxes \citep{Grete2021ApJ,Grete2023ApJ,Federrath2021NatAs,Beattie2024arXiv240516626B}.
 This use of large, fixed central grids  allows us to quantitatively study the evolution of the intracluster and intragroup medium with no artifacts resulting from changes in refinement.
In particular, we wish to examine the interactions between the AGN jets and their environmens, including the driving of turbulence and dissipation of energy, the amplification and structure of magnetic fields in groups and clusters, and the impact that these phenomena have on the formation of cold, dense gas that may, in turn, feed the supermassive black hole and power the AGN.

In this paper we focus on a subset of our calculations -- a suite of idealized galaxy clusters with masses and other properties comparable to the Perseus cool-core cluster, with AGN feedback triggered by the formation of cool gas in the vicinity of the central supermassive black hole -- and present key results from those calculations. Following papers will delve deeper into the the velocity structure functions of the hot and cold phases and projection effects \citep{Fournier2025A&A}, evolution of magnetized turbulence in these clusters, closer comparisons to observation, the formation and properties of cold, magnetized filaments in the center of the clusters, and of halos with a range of different virial masses and properties.

This paper is structured as follows:  In Section~\ref{sec:methods} we describe our code and numerical methods, simulation setup, and the details of our AGN triggering and feedback algorithms.  In Section~\ref{sec:results} we survey the key results from the simulation suite, and discuss those results in Section~\ref{sec:discussion}.  Finally, in Section~\ref{sec:conclusions} we summarize our results and discuss future work that will emerge from analysis of this simulation suite.

\section{Methods}\label{sec:methods}


In this section, we describe the \athenapk code
(Section~\ref{sec:athenapk}), the numerical methods that we use for these
simulations, including the prescription used for AGN triggering and feedback
(Section~\ref{sec:simulation_setup}), and summarize the entire
simulation suite (Section~\ref{sec:simulation_suite}).

\subsection{The \athenapk code}
\label{sec:athenapk}

\athenapk is an open source,  performance portable, finite volume (magneto)hydrodynamics
code based on the adaptive mesh refinement framework \parthenon \citep{Parthenon} --
originally derived from \athenapp \citep{stoneAthenaAdaptiveMesh2020}.
Performance portability refers to the capability to compile (and optimize) the code to
various architectures (such as CPUs with different architecture or GPUs from different
vendors) using a single code base and is realized by use of the \kokkos programming
model \citep{Trott2022}.
\athenapk has proven scalability to the largest scales possible today, e.g., $\gtrsim 93\%$
weak scaling up to $73{,}000$ GPUs on Frontier (TOP500 \#1 from 06/2022 to 06/2024).
Further details on the challenges encountered running large scale simulation are detailed
in Sec.~\ref{sec:disc-comp-approach}.

While \athenapk implements a variety of numerical methods, all simulations presented
in this paper employ an overall second-order accurate, shock-capturing, finite volume
scheme consisting of 
RK2 time integration, piecewise-linear reconstruction, and a HLLD (MHD) or HLLC (hydro)
Riemann solver \citep{Miyoshi2005}.
The resulting mass fluxes calculated from the Riemann solver are also used to advect an
arbitrary number of passive tracer fluids.
In the MHD case, the hyperbolic divergence cleaning method presented in \citet{Dedner2002}
handles the $\nabla \cdot \mathbf{B} = 0$ constraint.
Optically thin cooling is calculated by the extract integration method introduced by
\citet{Townsend2009}.

For numerical stability -- especially in cases where the fast, hot jet interacts with
cold clumps, first order flux correction is used.
All fluxes in cells for which a normal update would yield a negative pressure or density
are recalculated using a more diffusive scheme consisting of piecewise constant reconstruction
and an LLF Riemann solver.

Simulations were conducted using commit $\texttt{3ce0a88}$ and the complete input file for
the \fid simulation is available online as supplemental material.

\subsection{Simulation Setup}
\label{sec:simulation_setup}

The simulations are run on a Cartesian grid in a cubic volume with a side length of $6.4~\text{Mpc}$, covered by $1{,}024^3$ cells in the
base grid in a hierarchy of static meshes.
We enforce $3$ levels of refinement with $[-400,400]^3~\text{kpc}$ (where the root grid is the 0th level),
$5$ levels of refinement on $[-200,200]^3~\text{kpc}$,
and $6$ levels of refinement on $[-125,125]^3~\text{kpc}$.
Thus, the central region of the simulation is covered with a uniform grid of $2{,}560^3$ cells with cell side length of $\Delta_x \approx 100$\,pc.

Cosmological expansion is neglected in these simulations.  We
used a vanilla {\textLambda}CDM model to get the virial mass of the NFW halo and to
set its gas temperature. We set redshift $z=0$ at initialization with $\Omega_M = 0.3$, $\Omega_\Lambda = 0.7$, and $H_0 = 70 \text{ km s}^{-1}$.  We note that the precise details of the cosmological model do not impact the results presented in later sections of this paper, which pertain to baryonic physics in the halo core. 

The specific values given along the individual method descriptions pertain to the Perseus-like cluster setup.

\subsubsection{Gravitational Potential}

The gravitational potential has three components: a dark matter halo
profile, a BCG with a mass profile, and a SMBH. 
The dark matter follows the NFW profile \citep{navarroUniversalDensityProfile1997},
using $M_{\rm NFW} = 6.6 \times 10^{14}~\text{M}_\odot$ for the mass of the halo
\citep[corresponding to $M_{200}$, i.e, the mass contained within the virial radius $r_{200}$ within which the mean
enclosed mass density is 200 times the critical density of the
Universe at the cluster redshift;][]{Simionescu2011}
and a concentration parameter $c_{\rm NFW} = 6$.
The gravitational field from the NFW profile takes the form
\begin{equation}
  g_{\text{NFW}}(r) = 
    \frac{G}{r^2} \frac{M_{\rm NFW}  \left [ \ln{\left(1 + \frac{r}{R_{\rm NFW}} \right )} - \frac{r}{r+R_{\rm NFW}} \right ]}{ \ln{\left(1 + c_{\rm NFW}\right)} - \frac{ c_{\rm NFW}}{1 + c_{\rm NFW}} }.
\end{equation}

The scale radius, $R_{\rm NFW}$, for the NFW profile is computed from
\begin{equation}
R_{\rm NFW} = \left ( \frac{M_{\rm NFW}}{ 4 \pi \rho_{\rm NFW} \left [ \ln{\left ( 1 + c_{\rm NFW} \right )} - \frac{c_{\rm NFW}}{1 + c_{\rm NFW}} \right ] }\right )^{1/3} ,
\end{equation}
where the scale density $\rho_{\rm NFW}$ is computed from 
\begin{equation}
\rho_{\rm NFW} = \frac{200}{3} \rho_{crit} \frac{c_{\rm NFW}^3}{\ln{\left ( 1 + c_{\rm NFW} \right )} - c_{\rm NFW}/\left(1 + c_{\rm NFW} \right )}.
\end{equation}
The critical density, $\rho_{crit}$, is defined as:
\begin{equation}
    \rho_{crit} = \frac{3 H_0^2}{8 \pi G}.
\end{equation}

We use a Hernquist BCG profile
\begin{equation}
    g_{\rm BCG}(r) = G \frac{ M_{\rm BCG} }{R^2} \frac{1}{\left( 1 + \frac{r}{R}\right)^2}
\end{equation}
with $M_{\rm BCG} = 2.4 \times 10^{11}~M_\odot$  and $R_{\rm BCG} = 10~\text{kpc}$
\citep{Mathews2006}.

We include the gravitational field from a SMBH black hole with $M_{\rm
  SMBH} = 1.1 \times 10^9~M_\odot$ at the center of the cluster halo
\citep{Riffel2020}, which is coincident with the center of both the dark
matter and BCG halos.

\subsubsection{Entropy Profile}
We initialize the specific entropy as a function of radius, where entropy $K$ is defined as 
\begin{equation}
  K \equiv \frac{ k_b T}{n_e^{2/3} }
\end{equation}
where $k_b$ is Boltzmann's constant, $T$ is the temperature, and $n_e$ is the electron density. We set the initial entropy profile of the gas as a power law plus a constant floor using the same form as the ACCEPT database
\citep{cavagnoloIntraclusterMediumEntropy2009}
\begin{equation}
  K(r) = K_{0} + K_{100} \left ( r/ 100 \text{ kpc} \right )^{\alpha_K},
\end{equation}
where $K(r)$ is the specific entropy as a function of radius and $K_0$, $K_{100}$, and $\alpha_K$ are parameters. We use $K_0=20.0 ~\text{keV cm}^2 , K_{100}=110.0 ~\text{keV cm}^2$, and $\alpha_K=1.1$ for the initial entropy profile of a Perseus-like cluster.

\subsubsection{Initial Pressure and Density (Hydrostatic Equilibrium)}
We compute the initial pressure and density by enforcing the initial cluster to be in hydrostatic equilibrium given the gravitational profile described above and the ACCEPT-like entropy profile, assuming an ideal gas with adiabatic index $\gamma=5/3$. 
In order to close the set of equations to define the initial gas profile, we fix the gas density at $r=1.8\text{ Mpc}$ to $\rho =9.47 \times 10^{-29} \text{ g cm}^{-3}$.

\subsubsection{Initial velocity and magnetic field perturbations}
\label{sec:initialV_B}

In order to break symmetry in the initial conditions and seed a
weakly turbulent background state, we initialize both
the velocity and magnetic field with perturbations.
They are generated in spectral space, each based on 40 wavemodes chosen
randomly within an interval of characteristic scales between 50 and 200\,kpc.
The amplitudes are set by an inverse parabolic shape with a peak at a characteristic
lengthscale of 100\,kpc and scaled to to a root mean squared velocity of
75\,km/s and magnetic field of $1\,\mu$G.

\subsubsection{Cooling and plasma composition}

We use a helium mass fraction $\chi = 0.25$, with the remaining baryonic mass being hydrogen and electrons, which allows temperature $T$ to be defined from density $\rho$ and pressure $P$ following
\begin{equation}
    T = \frac{\mu m_h}{k_B}\frac{P}{\rho}.
\end{equation}
where $m_h$ is the atomic mass of hydrogen, $k_B$ is Boltzmann's constant, and $\mu$ is the mean particle mass per $m_h$, found by
\begin{equation}
    \mu = \left [ \frac{3}{4} \chi + 2 \left ( 1 - \chi \right ) \right ]^{-1}.
\end{equation}

The plasma cooling rate (i.e., radiative loss rate) is based on
tabulated tables tables from \citet{schureNewRadiativeCooling2009a}
assuming 1 Z$_\odot$ metallicity for all our runs except for our
``weak cooling'' simulation where we assume a metallicity of 0.3
Z$_\odot$. While most previous work assumes one-third Solar metallicity for their fiducial run \citep[e.g.,][]{Li_2014,Wang_2021}, observations of cool-core clusters conclude that the metallicity of the ICM in the core region is usually closer to 0.5 -- 1 Z$_\odot$, and decreases with increasing radius to reach roughly one-third Solar metallicity for $r \sim 100$~kpc \citep{Sanders_2007,McDonald_2019}. It is thus likely that our strong cooling simulations overestimate cooling for radii larger than a few tens of kpc, and that our weak cooling simulation underestimates cooling in the core region.

\subsubsection{AGN Feedback}

We include AGN feedback using thermal heating, kinetic jet, and magnetic tower models exploring different relative strengths. We divide the AGN feedback between the three channels following

\begin{equation}
    \dot{E}_{\rm AGN} = \dot{E}_T + \dot{E}_K + \dot{E}_B = \left ( f_T + f_K + f_B \right )\dot{E}_{\rm AGN}.
\end{equation}
where $\dot{E}_{\rm AGN}$ is the total AGN feedback rate; $\dot{E}_T$, $\dot{E}_K$, and $\dot{E}_B$ are the total thermal, kinetic, and magnetic AGN feedback rates; and $f_T$, $f_K$, and $f_B$ are the thermal, kinetic, and magnetic fractions of the total AGN feedback rate.
Given the scales in our simulations we set $f_T = 0.25$, $f_K = 0.74$
and $f_B =0.01$ (or $f_K = 0.75$ and $f_B = 0$ in the hydrodynamic case).
The thermal and kinetic fractions are motivated by \citet{Meece2017ApJ} and the magnetic fraction
was chosen to be subdominant given the distance to the SMBH's accretion disk.  The value of $f_B$
is in agreement with general 
relativistic magnetohydrodynamical simulations of black hole-launched jets   
\citep[see, e.g.,][]{Liska2020MNRAS,Kaaz2023ApJ}, where the jet is magnetically-driven
at a few gravitational radii but is almost entirely kinetic at tens of thousands of
gravitational radii.

\begin{figure}[htbp!]
  \centering
  \includegraphics{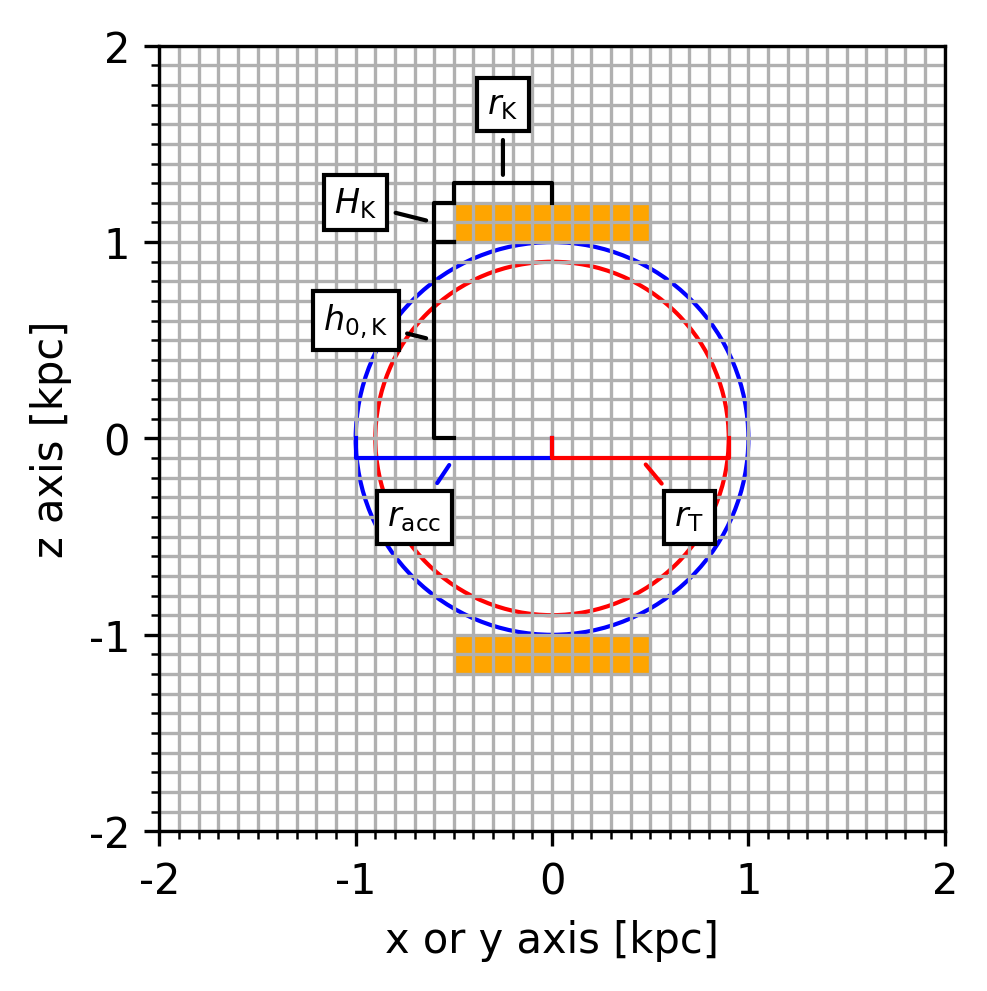}
  \caption{Sketch of some of the key model parameters and their
    values given the mesh setup in our simulations:
    the accretion radius, $r_\mathrm{acc}$,
    the thermal feedback radius, $r_\mathrm{T}$ (set to a smaller
    value here for illustrative purposes 
    whereas in the simulations $r_\mathrm{T} = r_\mathrm{acc}$),
    the kinetic jet launching radius $r_\mathrm{K}$, height $h_\mathrm{0,K}$, and offset $h_\mathrm{K}$.
    Small squares correspond to a single cell in the simulation.
  }
  \label{fig:model-sketch}
\end{figure}
A sketch of some of the key model parameters and their values used for the simulations 
is shown in Fig.~\ref{fig:model-sketch}.

\subsubsection{Thermal AGN Feedback}
In the thermal feedback model, thermal energy is deposited volumetrically within a sphere around the center of the halo where the presumed AGN resides~\citep{Booth2009}.
\begin{equation}
    \dot{e}_T \left ( r \right )  =  
    \begin{cases}
    \frac{f_T \dot{E}_{\rm AGN}}
    {\left ( 4/3 \right) \pi R_T^3 }, & \mathrm{if}~r \leq R_T, \\
    0, & \text{otherwise,}
\end{cases}
\end{equation}
where we use $R_T=1\text{ kpc}$ for the radius of thermal feedback. We also inject mass proportional to the thermal feedback at uniform density throughout the sphere of thermal injection
\begin{equation}
    \dot{\rho}_T \left ( r \right )  =  
    \begin{cases}
    \frac{f_T \dot{M}_{\rm AGN}}{ \frac{4}{3} \pi R_T^3}, & \mathrm{if}~r \leq R_T, \\
    0, & \text{otherwise,}
\end{cases}
\end{equation}
so that the total mass injected will match
\begin{equation}
    \dot{M}_T = f_T \dot{M}_{\rm AGN} .
\end{equation}

\subsubsection{Kinetic AGN Feedback}

In the kinetic feedback model, kinetic energy and mass is injected into two disks offset above and below the midplane of the AGN jet. These disks have a radius of $R_K$, each with thickness $H_K$, and are offset from the midplane by $h_{0,K}$ such that kinetic feedback is injected as far away from the midplane as $H_K + h_{0,K}$. Following \citet{Meece2017ApJ}, the rate of mass-energy injected by the jet is set proportional the fraction of accreted mass afforded to kinetic feedback, with $\epsilon$ fraction of that mass-energy proportioned to energy, i,.e.,
\begin{align}
    f_K \dot{M}_{acc} c^2 &= \dot{M}_K c^2 + \dot{E}_K \\
    \dot{M}_K & = \left( 1 - \epsilon \right) \dot{M}_{acc} \\
    \dot{E}_K & = \epsilon \dot{M}_{acc} c^2 ,
\end{align}
where $c$ is the speed of light.
The jet then injects a mass density at the rate of 
\begin{equation}
    \dot{\rho}_{K} = \frac{\dot{M}_{K}}{2 \pi R_{K}^2 H_K} ,
\end{equation}
where $R_K=500$ pc and $H_K = 2 \Delta_x \approx  200$ pc
-- chosen so that the jet and associated physical instabilities at
the boundary layer are well resolved , cf., \citet{Weinberger2023MNRAS}.
The jet speed of the injected material is given by:
\begin{equation}
    v_{\text{K}} = \sqrt{2 \left( \eta \, c^2 - (1 - \eta) \, u_{\text{jet}} \right)} ,
\end{equation}
where $\eta = 10^{-3}$ is the feedback power efficiency and $u_{\text{jet}}$ the specific internal energy of the jet
(here set by a temperature $T_\mathrm{jet}=10^8$\,K).
The momentum of the injected material is aligned with the $z$-axis of the Cartesian coordinate system.\footnote{
The implementation of the kinetic feedback model also supports precessing jets, which
have not been used in the simulations presented in this paper.
}
The momentum density injected is then given by:
\begin{equation}
    \dot{\mathbf{M}}_K \left ( \mathbf{r} \right )  =  
    \begin{cases}
    \text{sign}(z)
    \dot{\rho}_K  v_K \hat{h}, & \parbox{0.4\linewidth}{if  $r \leq R_{K}$  and 
      $h_{0,K}~\leq~|h|~\leq~h_{0,K}~+~H_{K}$} \\
    0, & \text{otherwise,} 
    \end{cases}
\end{equation}
where $r$ here is the distance from the jet axis and $h$ is the signed height above or below the mid plane. The injected kinetic energy rate per volume is 
\begin{equation}
    \dot{\mathbf{e}}_K \left ( \mathbf{r} \right )  =  
    \begin{cases}
    \frac{1}{2} \dot{\rho}_K  v_K^2, & \text{if } r \leq R_{K} \text{ and } |z| \leq H_{K}, \\
    0, & \text{otherwise,} 
    \end{cases}
\end{equation}
so that the total kinetic energy injected matches $f_K E_{\rm AGN}$.

\subsubsection{Magnetic AGN Feedback}

In general, the magnetic AGN feedback implementation consists of two key pieces: the
magnetic field configuration and its normalization to match the target power.

For the magnetic field configuration, we use a simple closed field loop model --
effectively a magnetic donut -- given by the following vector potential
\begin{equation}
  A_h(r, \theta, h) = 
  \begin{cases}
    B_0 L_\mathrm{M} \exp{\left ( -r^2/L_\mathrm{M}^2 \right)}, & \parbox{0.3\linewidth}{if
    $h_\mathrm{0,M} \leq |h| \leq h_\mathrm{0,M} + h_\mathrm{M}$, } \\
    0, & \mathrm{otherwise,} 
    \end{cases}
    \end{equation}
    where $B_0$ is a to be determined normalization factor, $L_\mathrm{M}$ a characteristic
    length scale, and $h_\mathrm{0,M}$ and $h_M$ the injection offset in vertical direction
    and injection height, following the kinetic parameters.
Using the vector potential as basis allows the resulting magnetic field
\begin{equation}
  B_\theta(r, \theta, h) = 
  \begin{cases}
    2 B_0 r /L_M \exp{\left ( -r^2/L_M^2 \right)}, & 
    \parbox{0.3\linewidth}{if $h_\mathrm{0,M} \leq |h| \leq h_\mathrm{M} + h_\mathrm{0,M}$} \\
    0, & \mathrm{otherwise}.
  \end{cases}
\end{equation}
to be divergence-free to machine precision.
For all simulations the scales are chosen\footnote{
A more complex pinching magnetic tower model, as well as mass injection associated with
the magnetic feedback channel, are also implemented in the code but not used in the 
simulations presented in this paper given their large-scale focus.
}
such that the magnetic donut is effectively seeded within the kinetic jet launching region, i.e., $h_\mathrm{0,M} = h_\mathrm{0,K} = 1$\,kpc,
$h_\mathrm{M} = h_\mathrm{K} = 2 \Delta_x \approx 200$\,pc, and $L_M =250$\,pc, similar to \citet{Weinberger2017}.

Given that the injected magnetic field strength should be normalized with respect
to a target power rather than a fixed strength we apply the following steps:

 To inject a magnetic field of strength $B_0$, we set the initial magnetic field to ${\mathbf{B} = \mathbf{\mathcal{B}}|_{\mathcal{B}_0 = B_0}}$ with the vector field $\mathbf{\mathcal{B}}$ corresponding
to the magnetic field configuration. To inject a field with a rate $\dot{B}_0$, we add a magnetic field ${\dot{\mathbf{B}} = \mathbf{\mathcal{B}}|_{\mathcal{B}_0 = \dot{B}_0}}$ to the existing fields.
Injecting by magnetic power, however,  requires extra steps since the increase in magnetic energy depends on the existing magnetic field.  Given an existing magnetic field $\mathbf{B}_n$, injecting a magnetic field of strength $B_{\rm p}$ (which we need to solve for) leads to a new magnetic field strength

\begin{equation}
\mathbf{B}_{n+1} = \mathbf{B}_n +  \mathbf{\mathcal{B}}|_{\mathcal{B}_0 = B_{\rm p} }.
\end{equation}
The change in total magnetic energy is then
\begin{align}
    \label{eq:magnetic_tower_reductions}
    \Delta E_B  
                =& B_{\rm p} \int_\Omega  \mathbf{B}_n \cdot \mathbf{\mathcal{B}}|_{\mathcal{B}_0 = 1} \, dV
                 \\ &+ B_{\rm p}^2 \int_\Omega \frac{1}{2} \mathbf{\mathcal{B}}|_{\mathcal{B}_0 = 1} \cdot \mathbf{\mathcal{B}}|_{\mathcal{B}_0 = 1}
                 \, dV ,
\end{align}
where $\Omega$ is the simulation domain.
To determine the factor $B_{\rm p}$ to give the correct increase in magnetic energy, the two integrals in Eq. \ref{eq:magnetic_tower_reductions} corresponding to the linear and quadratic contributions must first be computed (via reduction over the entire domain). Then $B_{\rm p}$ can be determined by the quadratic formula where only one root will be positive. 
For the case of magnetic field injection by the AGN, the change in magnetic energy is set to $\Delta E_B = \Delta t f_B \dot{E}_{\rm AGN}$ and $B_{\rm p}$ is determined by the reductions above.

\subsubsection{AGN cold mass triggering}

In our simulations, AGN feedback is triggered by cold mass around the presumed AGN. AGN triggering occurs within a $r_{\rm acc}=1\text{ kpc}$ radius accretion zone around the presumed AGN.
Within the accretion zone, gas with a temperature below the threshold $T_{\rm cold}=5 \times 10^4 \text{ K}$ triggers AGN feedback. The mass accretion rate onto the AGN follows
\begin{equation}
    \dot{M}_{\rm AGN}  =  \int_{r<r_{\rm acc} }
    \rho_{\rm cold}(\mathbf{r})/t_{\rm acc}
    \text{d}V ,
\end{equation}
where $\rho_{\rm cold}(\mathbf{r})$ is equal to $\rho(\mathbf{r})$ in cells where $T(\mathbf{r}) \leq T_{\rm cold}$ and $0$ otherwise, and $t_{\rm acc}=100 \text{ Myr}$ is the accretion time scale. The total AGN feedback rate is then set to 
\begin{equation}
    \dot{E}_{\rm AGN} = \epsilon_{\rm AGN} \dot{M}_{\rm AGN} c^2 ,
\end{equation}
where $\epsilon_{\rm AGN}=10^{-3}$ is the efficiency with which rest-mass from the accreted cold gas is turned into feedback energy following, e.g., \citet{Gaspari2013MNRAS,Li_2014,Meece2017ApJ,Wang_2021}.

The accreted mass is removed from the simulation. Mass is only removed from cells within the accretion zone if they have a temperature below the cold gas temperature threshold. The density removed follows the rate
\begin{equation}
    \dot{\mathbf{\rho}} \left ( \mathbf{r} \right )  = 
    \begin{cases}
    \rho ( \mathbf{r} )/t_{acc},  & \text{if } T (\mathbf{r})< T_{cold}, \\
    0, & \text{otherwise.} 
    \end{cases}
\end{equation}
For model completeness we implemented other types of accretion, e.g., ``boosted Bondi,'' but these are not used in the present simulations.

\subsubsection{Stellar feedback}
\label{sec:stellar-feedback}

Heat and mass feedback from Type Ia supernovae (SNIa) are injected in a spherically symmetric kernel proportional to the (fixed) stellar mass density of the central cluster galaxy. Following \citet{Voit2015a,Prasad2020ApJ}, the energy density and mass density injected into a cell at distance $r$ from the cluster center is 
\begin{align}
    \dot{e}_{\rm SNIa} = \Gamma_{\rm SNIa} E_{\rm SNIa} \rho_{\rm BCG}(r) \\
    \dot{\rho}_{\rm SNIa} = \alpha \rho_{\rm BCG}(r) ,
\end{align}
where $\Gamma_{\rm SNIa}= 3 \times 10^{-14} \text{ SNIa}\text{ yr}^{-1}\text{ Msun}^{-1}$ is the SNIa rate, $E_{\rm SNIa} = 10^{51}\text{ erg} \text{ SNIa}^{-1}$ is the energy injected per SNIa, and $\alpha = 10^{-19} \text{ s}^{-1}$ is the mass injection rate by SNIa.

In addition to the SNIa channel we also include a second, instantaneous stellar feedback
channel to account for the absence of separate  star particles in the simulation.
They are typically included as mediators to convert high density gas that would end up in
stars into thermal energy.
Therefore, a fraction of the gas above a number density threshold of $n > 50\,\mathrm{cm}^{-3}$,
a temperature below $T < 2 \times 10^4$\,K, and
within a radius of $r_\mathrm{acc} < r < 25$\,kpc is locally converted into
heat with an efficiency of $5 \times 10^{-6}$.

\subsubsection{Under the rug: on floors and ceilings}

Given the extreme conditions in the central region of the simulation, we limit certain
quantities to remain within the reasonable limits of the overall equation system solved
(e.g., not entering a relativistic regime).
More specifically, within $r < 20$\,kpc, the velocity is limited to $0.05c$, the
Alfv\'en velocity is also limited to $0.05c$, and the temperature limited to $5\times 10^9$\,K,
which are implemented as ceilings following the feedback routines.
To measure the impact of these limits we tracked the amount of energy being removed from the simulation.
Their effective powers were significantly smaller ($\ll 10^3$) than the AGN feedback and, thus, negligible
for the dynamics in the simulation.
In addition, we employ a global temperature floor at $10^4$\,K, which is below the lower
end of the cooling tables.

\subsection{The simulation suite}
\label{sec:simulation_suite}

\begin{figure*}[htbp!]
  \centering
  \includegraphics{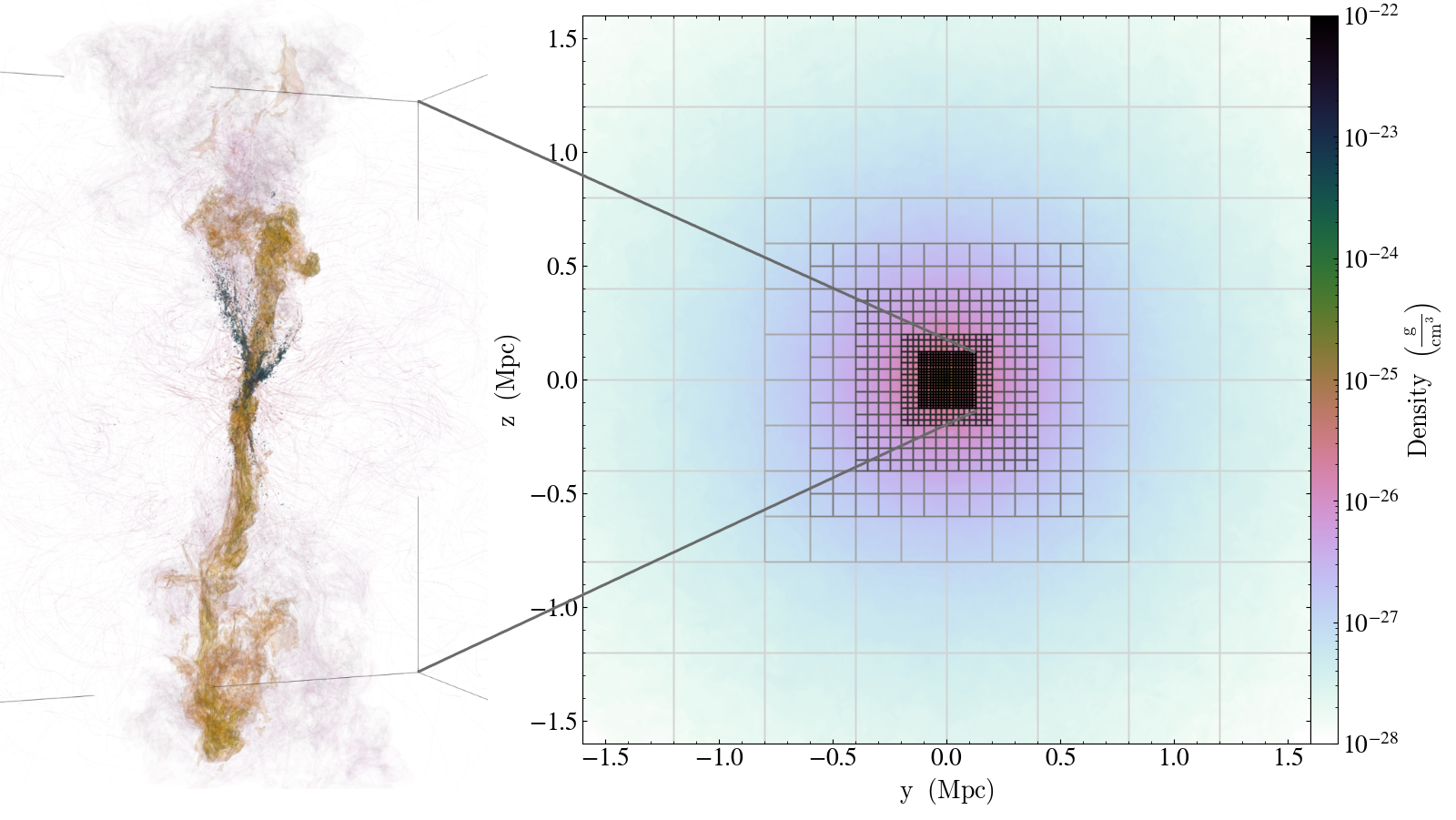}
  \caption{Illustration of the mesh (excluding the root mesh covering a [$6.4$\,Mpc]$^3$ domain) used in the cluster simulations.
    Each square is a cube containing $128^3$ cells for a total of $\approx36$ billion cells.
    Overall, 6 levels of refinement are used so that
    the innermost $[250\,kpc]^3$ is covered by a uniform $2560^3$ 
    mesh with $\Delta_x = 100\,pc$.
    The volume rendering of the central region highlights jet material in orange colors and cold gas in dark blue.
  }
  \label{fig:mesh}
\end{figure*}

A systematic overview of the initial conditions of all current simulations in the 
XMAGNET suite are given in Table~\ref{tab:simparams}.
All the cluster simulations, i.e., Perseus-like, a larger one ($\mathtt{Lg}$), and a smaller
one ($\mathtt{Sm}$), use a uniform $2{,}560^3$ mesh covering the central region, see Fig.~\ref{fig:mesh} for an illustration,
whereas in the
  group simulations, $\mathtt{SPG}$ and $\mathtt{MPG}$, only the 
  $[50\,\mathrm{kpc}]^3$  are covered by a $512^3$ mesh at the highest resolution of
  $\Delta_x \approx 100$\,pc in the center.
Similarly, the initial velocity perturbations in the group simulation have been scaled down
to 30\,km/s at 40\,kpc scales.

Note that only the Perseus-like cluster results are going to presented
in the following sections of this paper.
Results from the large and smaller cluster setups and from the galaxy group simulations will be presented in separate manuscripts.

\begin{table*}[ht!]
  \centering
\begin{centering}

\begin{tabular}{llllllllllll}
\toprule
 Sim.                     & $B_0$     & $Z$         & $K_0$                          & $K_{100}$                      & $\alpha_{K}$   & $r_\mathrm{ref}$   & $M_\mathrm{NFW}$     & $c_\mathrm{NFW}$   & $M_\mathrm{BCG}$     & $r_\mathrm{BCG}$   & $M_\mathrm{SMBH}$   \\
                          & [$\mu$G]  & [$Z_\odot$] & [$\rm{cm}^{2} \cdot \rm{keV}$] & [$\rm{cm}^{2} \cdot \rm{keV}$] &                & [$\rm{Mpc}$]       & [$\rm{M}_\odot$]     &                    & [$\rm{M}_\odot$]     & [$\rm{kpc}$]       & [$\rm{M}_\odot$]    \\

\midrule
 $\mathtt{Fiducial}$      & 1         & 1.0         & 20                             & 110                            & 1.1            & 1.8                & $6.6 \times 10^{14}$ & 5.0                & $2.4 \times 10^{11}$ & 10                 & $1.1 \times 10^{9}$ \\
 $\mathtt{WeakCool}$      & 1         & 0.3         & 20                             & 110                            & 1.1            & 1.8                & $6.6 \times 10^{14}$ & 5.0                & $2.4 \times 10^{11}$ & 10                 & $1.1 \times 10^{9}$ \\
 $\mathtt{WeakMag}$       & $10^{-3}$ & 1.0         & 20                             & 110                            & 1.1            & 1.8                & $6.6 \times 10^{14}$ & 5.0                & $2.4 \times 10^{11}$ & 10                 & $1.1 \times 10^{9}$ \\
 $\mathtt{Hydro}$         & --        & 1.0         & 20                             & 110                            & 1.1            & 1.8                & $6.6 \times 10^{14}$ & 5.0                & $2.4 \times 10^{11}$ & 10                 & $1.1 \times 10^{9}$ \\[0.5em]
 $\mathtt{FiducialLg}$    & 1         & 1.0         & 15                             & 110                            & 1.1            & 2.28               & $1.3 \times 10^{15}$ & 5.0                & $1 \times 10^{12}$   & 7                  & $7 \times 10^{9}$   \\
 $\mathtt{WeakMagLg}$     & $10^{-3}$ & 1.0         & 15                             & 110                            & 1.1            & 2.28               & $1.3 \times 10^{15}$ & 5.0                & $1 \times 10^{12}$   & 7                  & $7 \times 10^{9}$   \\
 $\mathtt{HydroLg}$       & --        & 1.0         & 15                             & 110                            & 1.1            & 2.28               & $1.3 \times 10^{15}$ & 5.0                & $1 \times 10^{12}$   & 7                  & $7 \times 10^{9}$   \\[0.5em]
 $\mathtt{FiducialSm}$    & 1         & 1.0         & 1.53                           & 150                            & 1.1            & 1.1                & $1.5 \times 10^{14}$ & 5.0                & $5 \times 10^{11}$   & 7                  & $5 \times 10^{9}$   \\
 $\mathtt{WeakMagSm}$     & $10^{-3}$ & 1.0         & 1.53                           & 150                            & 1.1            & 1.1                & $1.5 \times 10^{14}$ & 5.0                & $5 \times 10^{11}$   & 7                  & $5 \times 10^{9}$   \\
 $\mathtt{HydroSm}$       & --        & 1.0         & 1.53                           & 150                            & 1.1            & 1.1                & $1.5 \times 10^{14}$ & 5.0                & $5 \times 10^{11}$   & 7                  & $5 \times 10^{9}$   \\[0.5em]
 $\mathtt{FiducialMPG}$   & 1         & 1.0         & 1.3                            & 150                            & 1.05           & 2                  & $4.4 \times 10^{13}$ & 9.5                & $1.2 \times 10^{11}$ & 1.2                & $4.6 \times 10^{8}$ \\
 $\mathtt{HydroMPG}$      & --        & 1.0         & 1.3                            & 150                            & 1.05           & 2                  & $4.4 \times 10^{13}$ & 9.5                & $1.2 \times 10^{11}$ & 1.2                & $4.6 \times 10^{8}$ \\[0.5em]
 $\mathtt{FiducialSPG}$   & 1         & 1.0         & 1.5                            & 400                            & 1.05           & 1.8                & $4 \times 10^{13}$   & 7.5                & $2 \times 10^{11}$   & 1.6                & $2.6 \times 10^{9}$ \\
 $\mathtt{FiducialCcSPG}$ & 1         & 1.0         & 1.5                            & 200                            & 1.05           & 1.8                & $4 \times 10^{13}$   & 7.5                & $2 \times 10^{11}$   & 1.6                & $2.6 \times 10^{9}$ \\
 $\mathtt{HydroSPG}$      & --        & 1.0         & 1.5                            & 400                            & 1.05           & 1.8                & $4 \times 10^{13}$   & 7.5                & $2 \times 10^{11}$   & 1.6                & $2.6 \times 10^{9}$ \\
 $\mathtt{HydroCcSPG}$    & --        & 1.0         & 1.5                            & 200                            & 1.05           & 1.8                & $4 \times 10^{13}$   & 7.5                & $2 \times 10^{11}$   & 1.6                & $2.6 \times 10^{9}$ \\
\bottomrule
\end{tabular}
\end{centering}
\caption{Key differentiating parameters of all simulations:
  The initial magnetic field strength is the RMS value of $B_0$ on
  100\,kpc scales, $Z$ is the metallicity of the cooling table, $K_0$ and $K_{100}$ the specific
  entropies in the center and at $r=100$\,kpc, respectively, $\alpha_{K}$ is slope of the entropy profile, $r_\mathrm{ref}$ is the radius where the
  initial density of $\rho(r_\mathrm{ref}) = 9.47\times10^{-29}$\,g/cm$^3$ is fixed,
  $M_\mathrm{NFW}(=M_{200}$), $M_\mathrm{BCG}$, and $M_\mathrm{SMBH}$ are the dark matter halo, BCG, and
  SMBH masses, $c_\mathrm{NFW}$ is the concentration parameter, and $r_\mathrm{BCG}$ is the scale
radius of the Hernquist profile for the BCG.
}
\label{tab:simparams}
\end{table*}

\section{Results}\label{sec:results}

\subsection{Self-regulated feedback}

\begin{figure}[ht!]
\includegraphics[width=0.5\textwidth]{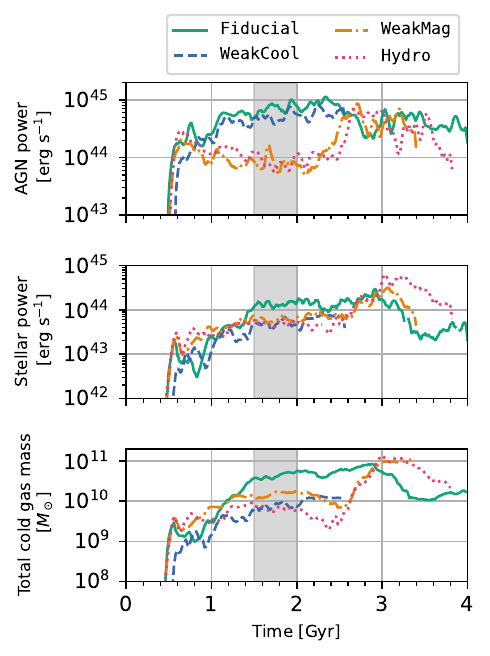}
\caption{From top to bottom: Temporal evolution of the AGN feedback power,
  instantaneous stellar feedback power (excluding Type Ia SNe), mass deposition rate, and total amount
  of cold gas.
The extend of the lines indicate the runtime of the simulations.
Given the 10x higher output frequency (every 2\,Myr) in the first two panels the data is smoothed over 50\,Myr using a box filter.
The shaded region illustrates the time interval that is used to quantity averaged properties in other figures.
}
\label{fig:power_vs_time}
\end{figure}

All of our simulations reach a state where the entire system self-regulates.
The top panel in Fig.~\ref{fig:power_vs_time} shows the AGN power over time.
In all cases it takes $\sim500$\,Myr before the AGN turns on for the first time.
In these first $\sim500$\,Myr the ICM plasma cools slowly with negligible amounts
of cold gas that are required to feed the AGN following the cold mass triggering model.

Once the AGN turns on, its power remains fairly steady at reasonable values between
$10^{44}$--$10^{45}$\,erg/s in all cases.
However, the \weakB and \hydro cases tend to be consistently lower at $\sim10^{44}$\,erg/s 
than the \fid and \weakC cases (with stronger initial fields) at $5\times10^{44}$\,erg/s.
Only at late times ($>2.5$\,Gyr) the power also increases in the \weakB and \hydro cases, which
can be traced back to AGN activity stimulating the formation of larger amounts
of cold gas, cf., the temporal evolution of the total cold ($T < 10^6$\,K) gas mass 
in the bottom panel of Fig.~\ref{fig:power_vs_time}.

In general, the amount of cold gas present in the cluster depends on 
both the magnetic
field and the cooling table, with an average cold gas mass between
1.5 and 2\,Gyr of
$2.8\times10^{10}\,\mathrm{M}_\odot$ (\fid),
$1.3\times10^{10}\,\mathrm{M}_\odot$ (\weakB),
$6.4\times10^{9}\,\mathrm{M}_\odot$ (\hydro), and
$5.5\times10^{9}\,\mathrm{M}_\odot$ (\weakC), respectively.
While \fid and \weakC contain the largest and smallest amounts of cold gas in the entire
simulation volume their AGN powers are comparable, implying that the additional
cold gas does not reach the accretion region.
The latter is likely related to the instantaneous stellar feedback that is converting
more cold gas into thermal energy in the \fid case, cf., the second panel in Fig.~\ref{fig:power_vs_time}.
The resulting densities in the launching cells is typically $\mathcal{O}(10^{-26})$\,g/cm$^{3}$.

\begin{figure*}[htb]
\centering
\includegraphics[width=\textwidth]{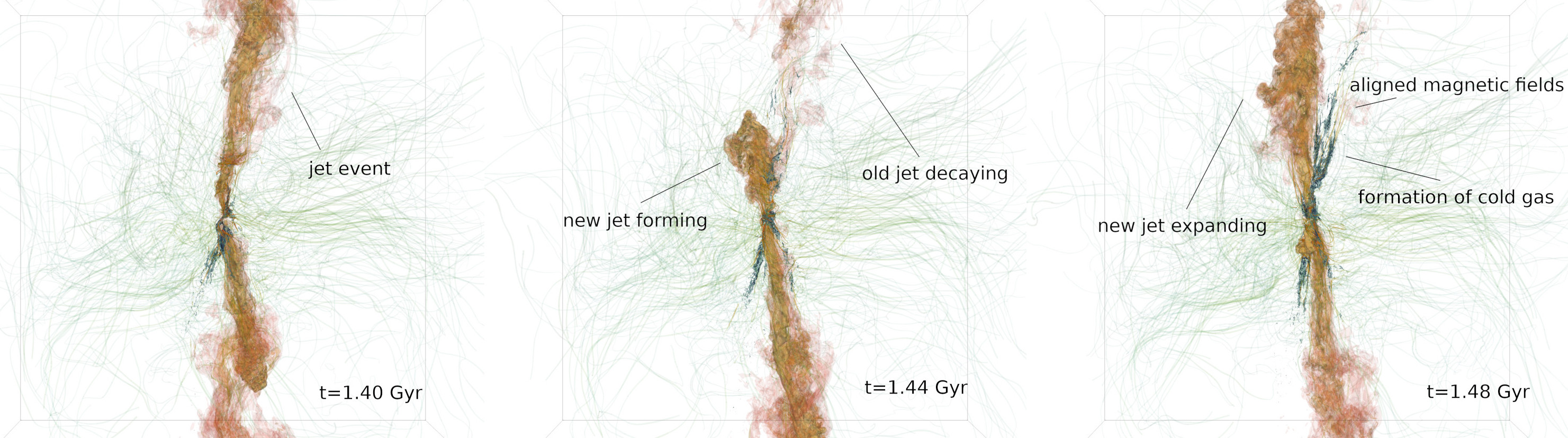}
\caption{Volume renderings of the central [$200$\,kpc]$^3$ in the \fid simulation 
  at $t=\{1.40,1.44,1.48\}$\,Gyr highlighting jet variability on short timescales
  (here 40\,Myr apart).
  Yellow colors indicate jet material (passive tracer concentration), dark blue colors
  show cold, $\approx 10^4$\,K, gas, and green lines illustrate magnetic fields.
}
\label{fig:evol_snaps}
\end{figure*}

Overall, the AGN power only exhibits limited variability and no clear ``duty cycles'' are
directly visible from the temporal evolution of the power itself.
However, individual outbursts are still clearly visible as illustrated in Fig.~\ref{fig:evol_snaps}.
The panels show three volume rendering of the passive tracer injected with the jet at snapshots
40\,Myr apart.
In the first panel, a jet is clearly visible extending in the top right direction.
40\,Myr later this jet event has been dissipated while another jet starts to form
towards the top left direction.
Finally, the last panel (another 40\,Myr later) illustrates the new
jet extending even further outwards.
It also shows the formation of cold gas following the direction
of the initial jet, highlighting the link between jet activity and the formation
of cold gas.

\begin{figure*}[htbp]
  \centering
  \includegraphics{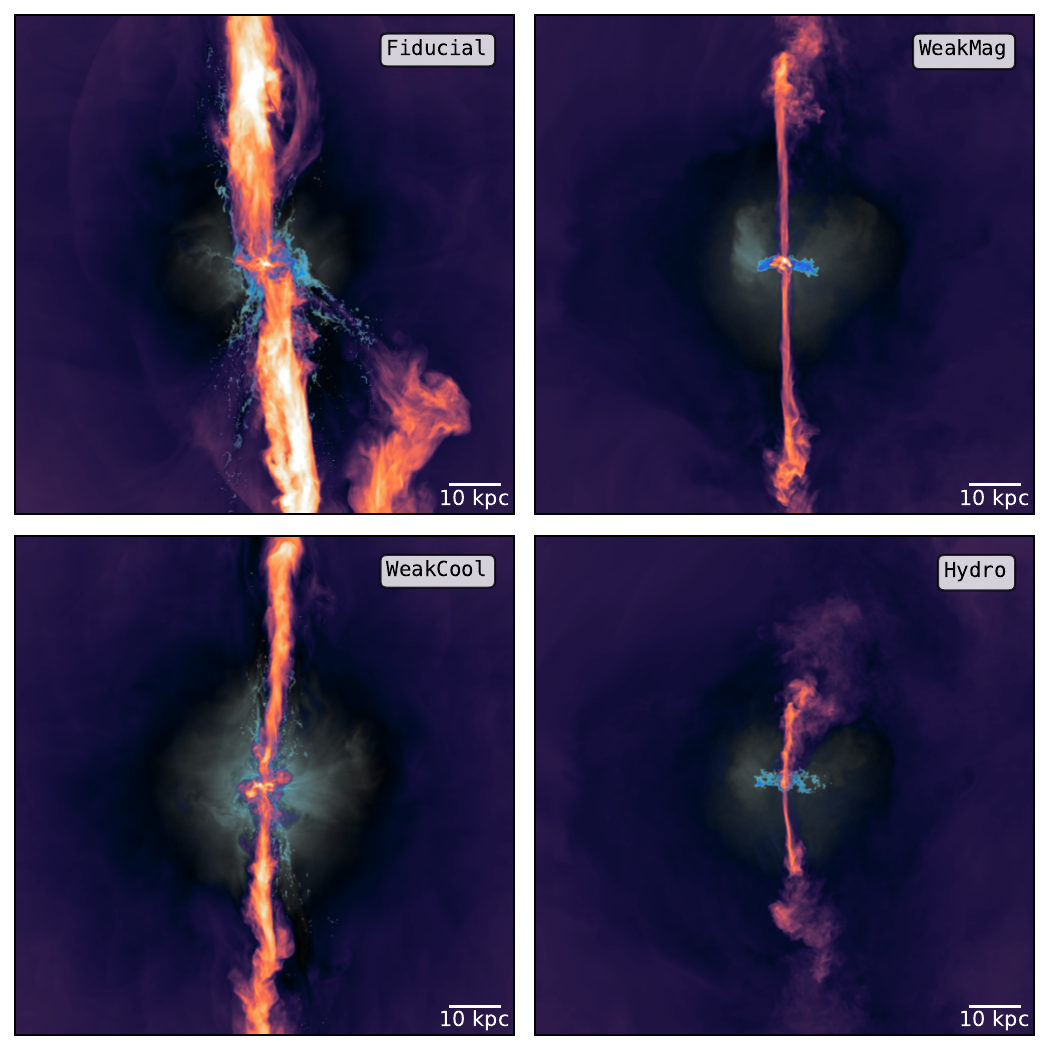}
  \caption{Projections with a depth of 10\,kpc of the central $([-50,50]\mathrm{kpc})^2$ at $t=1.72$\,Gyr highlighing the key dynamical differences between the different simulations (i.e.,the AGN power, and amount and distribution of cold gas). Dense gas is shown in bright blue colors and hot gas in red colors.
  }
  \label{fig:thin_proj}
\end{figure*}
For an overall impression of the general differences in morphologies between the simulations,
Fig.~\ref{fig:thin_proj} shows a 10\,kpc deep projection of both hot and dense gas.
\fid and \weakC both exhibit extended filamentary structures of cold gas.
Overall, the \fid case is more dynamic with more cold gas being available and visible at larger
distances from the center.
Similarly, the jet in the \fid case is visibly stronger and more dynamic than in the \weakC case.
\weakB and \hydro are in stark contrast to those dynamics.
In these latter simulations the cold gas is mostly confined to a disk around the center (indicating that weak magnetic
fields are not enough to prevent the formation of a disk as in the pure hydrodynamics case).
Moreover, the lower jet power in these cases is clearly visible and, in the absence of extended
cold gas to interact with, the jet itself is much narrower and more closely aligned with its
original launching direction.

\subsection{Thermodynamics and energetics}
\label{sec:energetics}
\begin{figure*}[!htb]
\centering
\includegraphics[width=\textwidth]{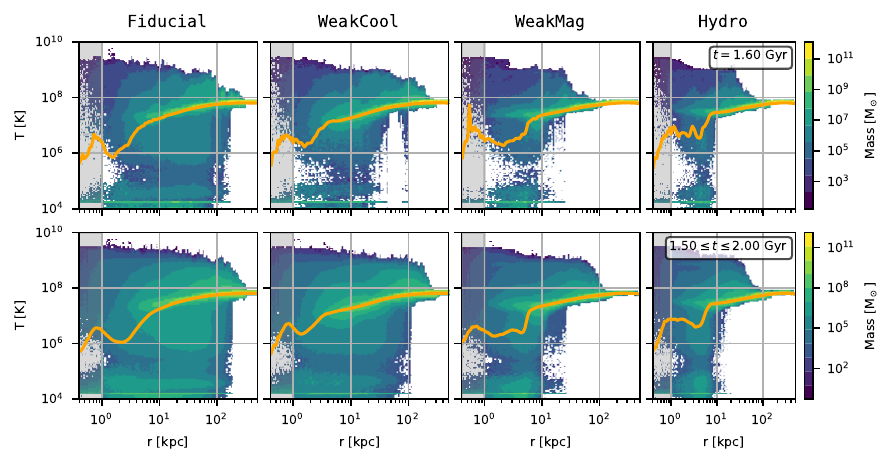}
\caption{Mass-weighted temperature histograms at $t=1.6$\,Gyr (top panels)
  and averaged between $1.5 \leq t \leq 2.0$\,Gyr (bottom panels) 
  of all simulations.
  The orange lines indicate the mass-weighted mean temperature versus radius.
  The gray shaded region illustrates the innermost kpc for which the statistics should be interpreted
  with care given the various active AGN model components.
}
\label{fig:T-vs-r}
\end{figure*}

The mass-weighted temperature histograms versus radius from the cluster center are shown in Fig.~\ref{fig:T-vs-r}.
Down to cluster-centric radius of 10\,kpc the average temperature does not differ much between setups and is
dominated by the bulk mass of the hot $10^7$--$10^{8}$\,K ICM.
The cold gas distribution, however, differs significantly between runs with (traces of)
cold gas visible up to 
$r\sim150$\,kpc in the \fid case, 
$r\sim100$\,kpc in the \weakC case, 
$r\sim30$\,kpc in the \weakB case, and
$r\sim20$\,kpc in the \hydro case.
This also applies to intermediate-temperature gas visible as vertical streaks extending
down from the hot, bulk ICM at different radii depending on the run.
Thus, both magnetic fields and the efficiency of the cooling impact the extent of cold gas with
the magnetic fields having a stronger effect -- though the impact may also be indirect, cf., the
varying AGN powers.
The overall behavior is equally visible in the time-averaged histograms over 500\,Myr shown
in the bottom panels of Fig.~\ref{fig:T-vs-r}.

\begin{figure}[htbp]
  \centering
  \includegraphics{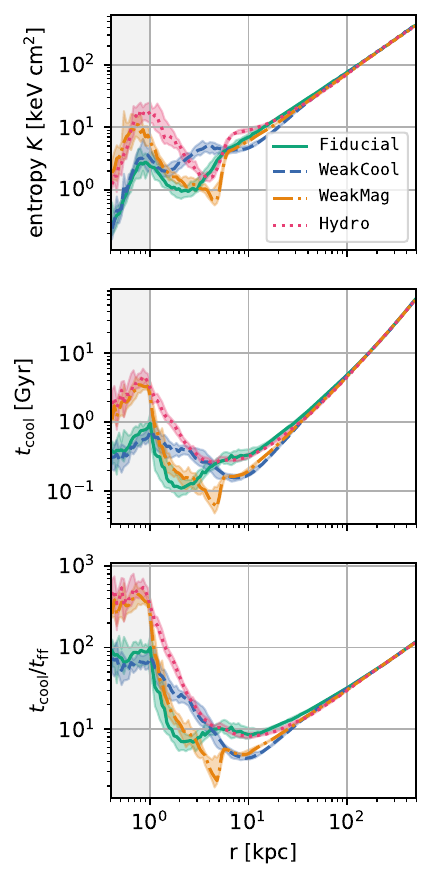}
  \caption{Median and inter quartile ranges (over time between $1.5 \leq t \leq 2.0$\,Gyr)
  of the mass-weighted mean entropy (top), cooling time (center), and cooling over freefall
time (bottom).
  Here and in subsequent plots, the gray-shaded region denotes the innermost kpc for which our results strongly depend on the AGN model.
  Plots show that min($t_{\rm cool}/t_{\rm ff}$) stays within $5 - 20$ for all the runs within $r < 50 $ kpc allowing for formation of extended cold gas filaments.
}
  \label{fig:Ktcool-r}
\end{figure}

From an even higher-level point of view, Fig.~\ref{fig:Ktcool-r} shows
that the median radial profiles over 500\,Myr of the entropy and
cooling time are indistinguishable between the runs down to 30\,kpc.
Within 30\,kpc the situation is more dynamic between the different runs with entropies of
order unity keV$\cdot$cm$^2$ and cooling times between 0.1--1\,Gyr.
No clear trend that differentiates between the metallicities of the cooling table nor
the initial magnetic field strength is visible between $1 \leq r \leq 10$\,kpc.

\begin{figure}[htbp]
  \centering
  \includegraphics{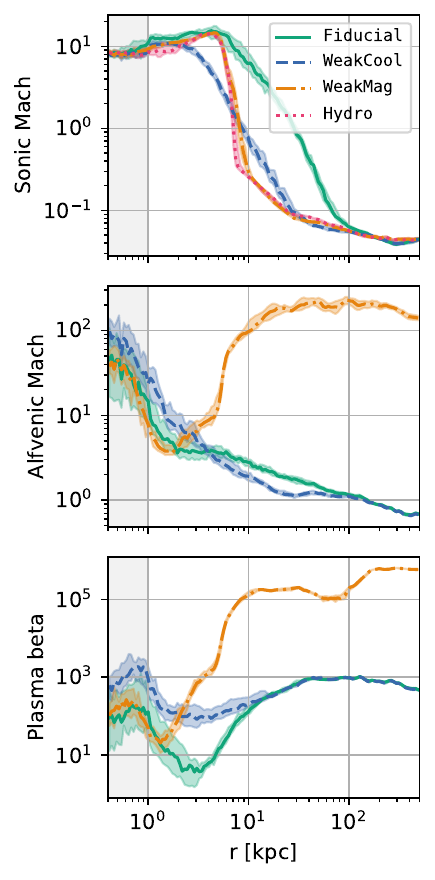}
  \caption{Median and inter-quartile ranges (over times $1.5 \leq t \leq 2.0$\,Gyr)
  of the mass-weighted mean sonic Mach number (top), Alfv\'enic Mach number (center), and plasma $\beta$ (bottom).
   Note the stark difference in the plasma  $\beta$ and Alfvenic Mach number for the weak magnetic field ($B=0.001\mu$ G) run compared to the fiducial run with $B=1\mu$ G indicates an effective magnetic field amplification owing to the $1/B^2$ scaling of the plasma $\beta$ and an initial difference in $B_0$ of $10^3$ between the \fid/\weakC and \weakB case.}
  \label{fig:MsMapb-r}
\end{figure}

The overall energy balance is depicted in Fig.~\ref{fig:MsMapb-r} based
on the median radial profiles of
the sonic Mach number (as a proxy for the kinetic energy over thermal pressure),
the Alfv\'enic Mach (as proxy for the kinetic over magnetic energy), and
plasma beta (as thermal pressure over magnetic energy/pressure).
The Alv\'enic Mach is comparable within $r\leq4$\,kpc (dropping from several tens to several few)
for all simulations that include magnetic fields.
Then it continues to decline, eventually becoming sub-Alfv\'enic 
beyond $r \gtrsim 100$\,kpc in the \fid and \weakC case whereas for \weakB it
increases again rapidly to $\sim200$, i.e., to a highly super-Alfv\'enic state.
This is also reflected in the plasma beta.
For \weakC it remains approximately constant within $r\leq500$\,kpc with values of several
hundreds up to 1000.
At $r>10$\,kpc this also applies to the \fid case. However, at small distances from the center
the plasma beta drops to $\mathcal{O}(10)$ in the \fid case and $\mathcal{O}(100)$ in
  the \weakC case, which is tied to the larger amont of cold
gas in the former compared to the latter.
Again, the \weakB simulation stands out with plasma betas reaching $10^5$ beyond $r>10$\,kpc.
Given the $1/B^2$ scaling of the plasma beta and an initial difference in $B_0$ of $10^3$
between the \fid/\weakC and \weakB case, this already gives a first indication of an effective
magnetic field amplification that is further discussed in Subsection~\ref{sec:magnetic-fields}.

\begin{figure*}[!htb]
\centering
\includegraphics[width=\textwidth]{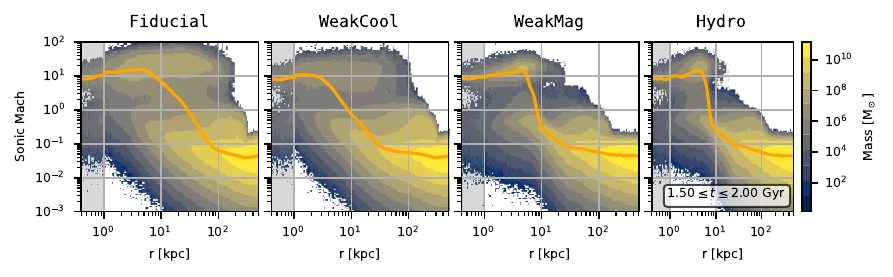}
\caption{Mass-weighted sonic Mach number histograms 
  averaged between $1.5 \leq t \leq 2.0$\,Gyr 
  of all simulations.
  The orange lines indicate the mass-weighted mean sonic Mach number versus radius.
}
\label{fig:Ms-vs-r}
\end{figure*}

Lastly, the radial profiles of the sonic Mach number shown in the top panel of Fig.~\ref{fig:MsMapb-r}
indicate that the innermost region $r\lesssim8$\,kpc is highly super-sonic with Mach
numbers reaching $\mathcal{O}(10)$, which then transitions to become subsonic
($\mathcal{O}(0.1)$).
In general, the \weakB and \hydro case are effectively indistinguishable with a sharp
transition between the super- and sub-sonic regime.
\fid and \weakC exhibit a smoother transition with the former only occurring at larger radii
($8 \lesssim r \lesssim 80$\, kpc versus $3\lesssim r \lesssim 30$\, kpc).
This is tied to the different cold gas distribution given the mass weighting in the radial profiles.
A more detailed picture is displayed in Fig.~\ref{fig:Ms-vs-r}, where the 2D histograms of
sonic Mach number versus radius are shown for all simulations.
An almost bimodal distribution is visible with a high (Mach $\sim10$) component
consisting of both the jet (fast and hot material) and cold gas (resulting in
a low sound speed).
This also explains why the sonic Mach number is more extended in the \fid case than for the other ones,
cf., the temperature profiles (and extent of cold gas) in Fig.~\ref{fig:T-vs-r}.
 
\subsection{Turbulent power spectra}

The large uniform resolution in the central region of our simulations allows us
to calculate the power spectra using standard methods typically applied in
idealized turbulent boxes.\footnote{We used the energy transfer analysis framework
  (\url{https://github.com/pgrete/energy-transfer-analysis}) to calculate the
  spectra \citep{Grete2017PhPl}, which assumes periodic boxes.
  Technically, a central cutout from our current simulations is not periodic.
  However, given the smooth, almost radial symmetric properties at
  $r \gtrsim100$\,kpc the cutouts are effectively symmetric introducing only very limited
noise close the grid cutoff scale.}
All spectra discussed cover the central $[200\,kpc]^3$ with a resolution of $2{,}048^3$
cells.

\begin{figure*}[htbp!]
  \centering
  \includegraphics[width=1.00\textwidth]{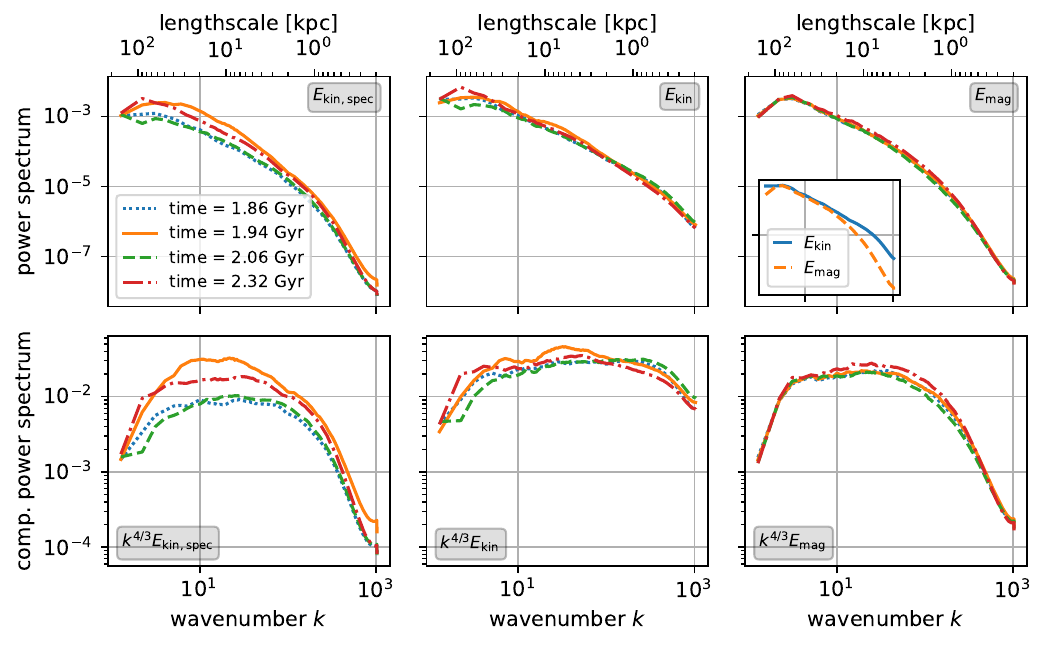}
  \caption{
    Instantaneous turbulent power spectra at various times of the specific kinetic 
    energy (left), the kinetic energy density (based on $\mathbf{w} = \sqrt{\rho}\mathbf{u}$, center panel),
    and magnetic energy density (right) in the \fid simulation.
    The raw, i.e., uncompensated spectra are shown in the top panels whereas the spectra in the bottom panels
    are compensated by $k^{1.2}$, $k^{1.1}$, and $k^{1.2}$, respectively.
    The inset in the top right panel shows the spectra of kinetic energy density and magnetic energy density 
    at $t=1.86$\,Gyr. The axes ranges in the inset are the same as in the larger panels.
  }
  \label{fig:spec-fid}
\end{figure*}
Fig.~\ref{fig:spec-fid} shows the turbulent specific kinetic energy, kinetic energy density, and magnetic
energy density power spectra at different times.
The times have been chosen to correspond to peak AGN power ($t=\{1.94, 2.32\}$\,Gyr) and low AGN
power in between ($t=\{1.86,2.06\}$\,Gyr).
It is immediate visible that the AGN is linked to the velocity fluctuations (i.e., the
specific kinetic energy power spectrum).
At higher AGN activity there is more power on all scales -- especially between
characteristic scales of $2 \lesssim \ell \lesssim  20$\,kpc.
At the two times with lower AGN activity (which surround the first peak time) the
spectra are effectively identical, highlighting the different phases of
AGN activity and their impact on the velocity structure.
Moreover, the activity also impacts the shape of the power spectra, cf., the bottom
left panel in Fig.~\ref{fig:spec-fid} which shows the compensated spectra and
for which a clear break in spectral slope is visible at $t=1.94$\,Gyr.
This potentially also applies to the kinetic energy density spectra, i.e., the ones based
on $\mathbf{w} = \sqrt{\rho}\mathbf{u}$ taking density fluctuations into account.
However, the differences in these spectra are much more subtle and barely visible.
Overall, the kinetic energy density spectra are effectively identical, as are the
magnetic energy density spectra (right panels of Fig.~\ref{fig:spec-fid}).
From a general energy balance point of view the inset in Fig.~\ref{fig:spec-fid}
compares kinetic and magnetic energy density spectra at the same time.
On effectively all scales (and particularly on small scales $\ell \lesssim 10$\,kpc)
the gas motion is dominated by velocity fluctuations.

\begin{figure*}[htbp!]
  \centering
  \includegraphics[width=1.00\textwidth]{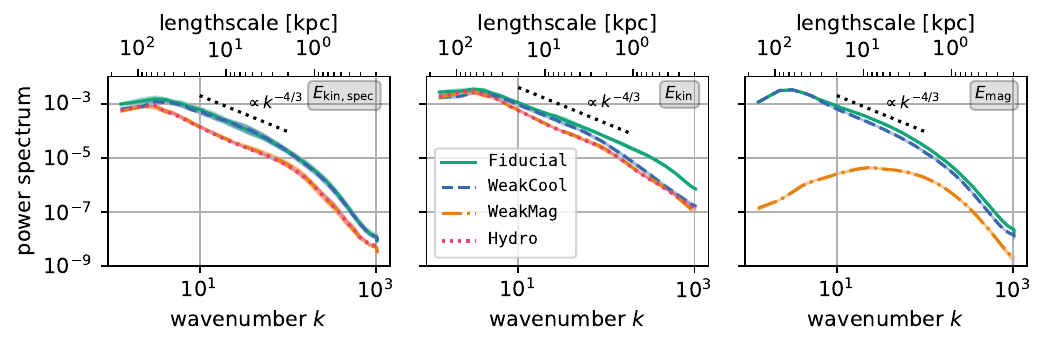}
  \caption{Median turbulent power spectra (with IQR in shaded regions) 
of the specific kinetic
    energy (left), the kinetic energy density (based on $\mathbf{w} = \sqrt{\rho}\mathbf{u}$, center panel),
    and magnetic energy density (right)
    between $1.5$--$2$\,Gyr for all four simulations.
    The slopes are just given for reference and do not represent physical models.
}
  \label{fig:spec-avg}
\end{figure*}
In additional to temporal variations, there also exist variation with respect
to the different simulation setups.
Fig.~\ref{fig:spec-avg} shows the median spectra (between $1.5$--$2$\,Gyr) of all four configurations
that exhibit extended regimes of power-law scaling.
The specific kinetic energy spectra (left panel) is effectively identical between runs with similar
initial magnetic field strength (and, thus, AGN power), i.e., \fid and \weakC, and \weakB and \hydro.
Similar to the instantaneous spectra, the latter two have less power on all scales.
Moreover, a hint of a break in the scaling is visible around $\ell \sim 10$\,kpc with a steeper
spectrum on larger scales.
The impact of the different cooling tables (and initial magnetic field strength) is clearly
visible in the kinetic energy density power spectra in the center panel of Fig.~\ref{fig:spec-avg}.
Again, \weakB and \hydro are indistinguishable from each other.
However, \fid and \weakC only follow each other down to scales of $\ell \gtrsim 5$\,kpc.
At those scales the spectrum of \weakC becomes significantly steeper whereas the \fid one
extends further, indicating an impact of the cold (high density) gas dynamics on small
scales.
Finally, the magnetic energy power spectrum in the right panel of Fig.~\ref{fig:spec-avg} is
again similar between the \fid and \weakC case, with slightly more power in the
magnetic field fluctuations on scales $\ell \lesssim 10$\,kpc in the \fid case with stronger
cooling.
The \weakB magnetic energy spectrum does not exhibit a power law scaling regime, peaks
around $\ell \sim 10$\,kpc, and the fluctuation generally remain weaker on all scales
compared to the other cases.
These differences are presented further in the following subsection.

\subsection{Magnetic fields}
\label{sec:magnetic-fields}
\begin{figure*}[htbp!]
  \centering
  \includegraphics{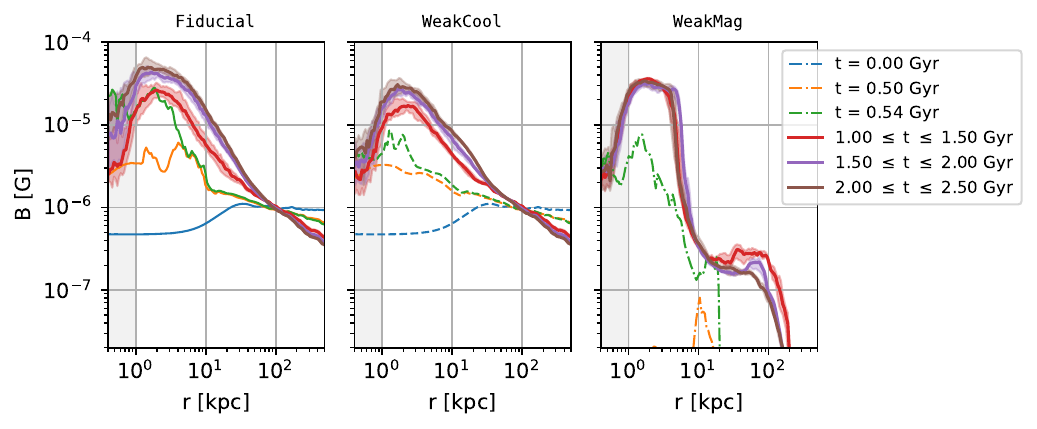}
  \caption{Mass-weighted magnetic field strength versus radius at various times and
    intervals (using the median and interquartile ranges as colored extents) for the
    \fid, \weakC, and \weakB simulation (from left to right).
}
  \label{fig:B-vs-r}
\end{figure*}
The radial profiles of the magnetic field strength at various times and intervals
are illustrated in Fig.~\ref{fig:B-vs-r}.
Independent of setup the magnetic fields reach strengths of several $10^{-5}$\,G within
$r\lesssim10$\,kpc even in the \weakB case and do not vary significantly after $1.5$\,Gyr.
At these late times, the fields towards larger radii in the \fid and \weakC cases continuously decrease smoothly
down to $10^{-5}$\,G at $r\sim100$,kpc whereas in the \weakB case they almost
follow two step functions down to a few $10^{-7}$\,G at $r\sim8$\,kpc and then another $>10\times$ decrease
around $r\sim100$\,kpc.
Beyond $r\gtrsim100$\,kpc the fields even decay from their initial values, making $r\sim100$\,kpc
a relevant scale out to which the AGN actively modifies the magnetic field configuration.
At smaller radii it takes several hundreds of Myr for for the profiles to stabilize even though
the initial magnetic field growth is fast.

\begin{figure*}[htbp!]
  \centering
  \includegraphics[width=1.00\textwidth]{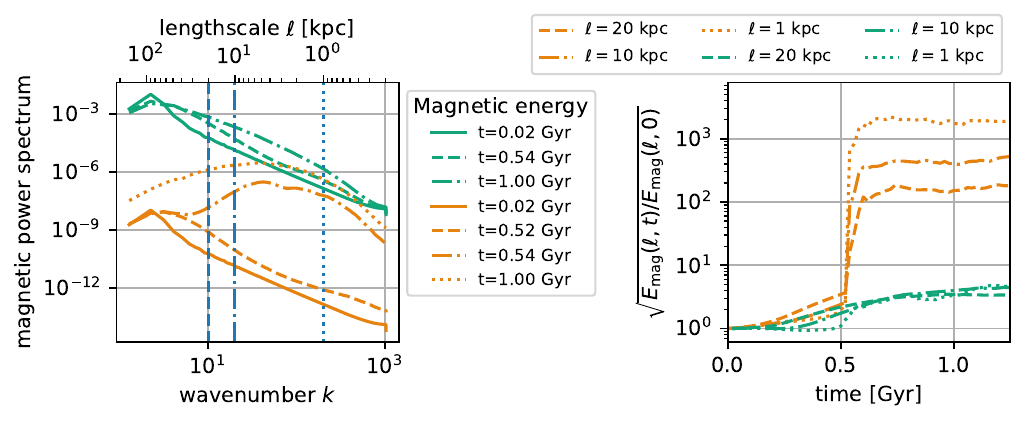}
  \caption{Left: Magnetic energy spectra of the \fid (green) and \weakB simulation (orange) at various times.
    The vertical blue lines indicate scales
  for which the power is plotted over time in the right panel (normalized to the initial value at the particular scale).}
  \label{fig:spec-mag}
\end{figure*}

This growth is further illustrated in Fig.~\ref{fig:spec-mag} where instantaneous
magnetic field power spectra are plotted around the first AGN outburst for the \fid and \weakB case
(left panel) along with the magnetic field amplification at scales of $\ell = \{1,10,20\}$\,kpc
for the first $1.25$\,Gyr.
For the stronger initial fields in \fid the amplification of the fields occurs gradually
at large scales (by a factor of $\sim4\times$ over $\sim500$\,Myr at $\ell = \{10,20\}$\,kpc) and 
faster on smaller scales ($4\times$ within $\sim150$\,Myr at $\ell = 1$\,kpc).
The magnetic field amplification in the \weakB case is much more extreme.
On all reference scales the magnetic field strength increases by $>100\times$ in
$\lesssim100$\,Myr and on the smallest scale ($\ell = 1$\,kpc) even by $\gtrsim100\times$
within just $20$\,Myr.
In the \weakB case the field growth continues, peaking at $\gtrsim1000\times$ on the small scales.

\subsection{AGN area of influence}
\begin{figure*}[!htb]
\centering
\includegraphics[width=\textwidth]{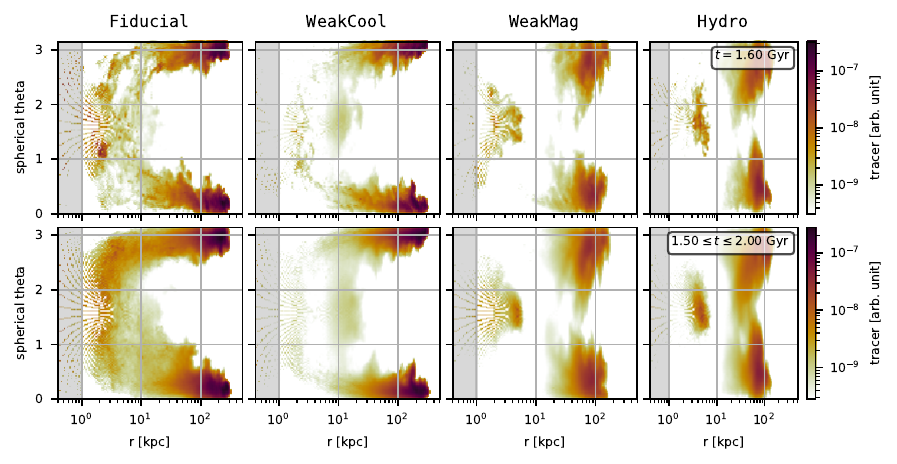}
\caption{Tracer concentration in spherical polar coordinate $\theta$ versus radius at $t=1.6$\,Gyr (top panels)
  and averaged between $1.5 \leq t \leq 2.0$\,Gyr (bottom panels) 
  of all simulations highlighting more concentrated jet material at higher altitudes in the
  \fid and \weakC cases, and a more isotropic distribution in the \weakB and \hydro cases.
}
\label{fig:theta-vs-r}
\end{figure*}

As presented in the preceding subsection, the AGN feedback impacts the magnetic fields out to $r\lesssim100$\,kpc, and, as shown in Subsection~\ref{sec:energetics},
the presence and/or formation of cold gas to various radii depends on the setup.
To understand where the plasma of the AGN jet ends up, the passive tracer concentration is shown in Fig.~\ref{fig:theta-vs-r} for spherical $\theta$ angle versus radius.

 AGN with higher powers (\fid and \weakC) push tracer material to higher altitudes
(up to $\sim300$\,kpc at 2\,Gyr compared to $\sim150$\,kpc for \weakB and \hydro).
Moreover, for \fid and \weakC the tracer concentration at radii larger than several tens of kpc is limited to low and
high values of $\theta$.
This means that the material is mostly concentrated around the jet axis. There
is no large-scale circulation through the midplane from larger radii.
In contrast, the \weakB and \hydro cases show a more isotropic distribution of the tracer material
both around the center (a few kpc scales) and around $70$--$100$\,kpc scales.

\section{Discussion}\label{sec:discussion}


\subsection{Cluster self-regulation}
\label{sec:disc-selfregulate}

The clusters discussed in Section~\ref{sec:results} all self-regulate, in the sense that the energy feedback from AGN and stars offsets the radiative cooling in the cluster core (defined for the purpose of this discussion to be the region within $\simeq 100$~kpc of the cluster center).  The feedback power lies between $10^{43}-10^{45}$ erg/s (as can be seen in Figure~\ref{fig:power_vs_time}).  
The simulations with high magnetic fields (i.e., consistent with observed cluster cores) typically have roughly a factor of $4-6$ higher sustained levels of AGN feedback than the low magnetic field or hydrodynamic (zero-field) simulations during the first $\simeq 2$~Gyr after the AGN starts. This corresponds to  times of roughly $0.4-2.4$~Gyr\footnote{We note that the hydrodynamical and weak initial field MHD runs abruptly form a large amount of cold gas at $t \sim 2.5$~Gyr, and hence we generally focus on evolution prior to that time.}.  In spite of significantly different AGN power during that time, the total amounts of cold gas in all four simulations are similar, even though the high cooling/high field ``fiducial'' run ultimately produces more cold gas than the other simulations.  Examination of Figure~\ref{fig:T-vs-r} suggests that this primarily caused by the distribution of cold gas rather than the total amount -- i.e.,  the strong field/strong cooling run has much more cold gas at large radii than the other simulations.

All of the AGN in our simulations operate continuously -- i.e., they do not have a discernable ``duty cycle'' where the AGN transitions between a state of high activity and one where it is either completely turned off or in a much lower state of activity.  This is possibly consistent with observations, which use a variety of means to estimate the duty cycle.  Observations by \citet{Vantyghem2014MNRAS} using primarily X-ray observations suggest a relatively low duty cycle, where the AGN is in its high state (i.e., where the AGN is injecting energy at rates comparable to the cluster's X-ray luminosity) significantly less than half of the time.  This result is in tension with a volume-limited X-ray sample of clusters by \citet{Panagoulia2014MNRAS}, who estimate a duty cycle of $\simeq 60\%$ for clusters with central cooling times of $\leq 3$~Gyr and $\gtrsim 80\%$ for clusters with a central cooling time of $\leq 0.5$~Gyr.  This latter result is in agreement with our simulated clusters, which all have central cooling times of $\leq 0.5$~Gyr in the central 20~kpc (central panel of Figure~\ref{fig:Ktcool-r}).  More recent radio observations of AGN jet-driven bubbles in the intracluster medium using LOFAR that include models of spectral aging for the electron synchrotron emission suggests that the AGN they observe has had a duty cycle close to unity in the past $\simeq 100$~Myr \citep{Biava2021A&A}.
These results are generally consistent with the behavior we see in all of our simulations.  We note, however, that the gas accretion timescale that we use ($T_{acc} = 100$~Myr) is relatively long and may smooth out accretion behavior.  Future simulations with varied accretion timescales will shed light on whether the high duty cycle that we observe is physical or numerical in nature.

The cluster entropy profiles show that all of the clusters are roughly consistent with X-ray observations, with central entropies in the $1-10$~keV range and central cooling times (within 10 kpc) in the $0.1-1$~Gyr ranges. The central entropy of the strong magnetic field runs tend to flatten toward the center, and the weak magnetic field and hydrodynamical runs tend to be higher near the center.  We infer that this has to do with the magnetic fields, but the causal mechanism is unclear.  The region within 1.0\,kpc of the cluster center is within the AGN accretion and thermal feedback regions, and thus any thermodynamic quantities that are measured should be interpreted with caution.
The difference in the entropy profile near the center between the strong magnetic field cases and the hydro and weak magnetic field cases is because there is more dense gas near the jet source region in those cases.  This can be see in Figure~\ref{fig:Ktcool-r}, where the hydro/weak field cases have higher $t_{\rm cool}$ within $r\lesssim 2$ kpc because that quantity depends inversely on density.

In terms of comparisons to prior simulation work relating to self-regulated clusters, we see that  in terms of bulk behavior our results are consistent with those of \citet{Meece2017ApJ}, who use a very similar triggering and jet injection mechanism, although with $2\times$ coarser spatial resolution (and effectively significantly worse effective spatial resolution given that they use the ZEUS hydro method, which uses artificial viscosity and has significant smoothing and mixing -- see \citet{OShea2005ApJS} for information). 

Our results differ from previous results of self-regulating galaxy clusters.  Hydrodynamic simulations of idealized clusters including cold gas-triggered AGN, such as those performed by \citet{Gaspari2012ApJ, Prasad2015ApJ, Li2015ApJ} and \citet{Meece2017ApJ},  typically show a greater variability in AGN energies and large-scale cyclic behavior on $\sim$ Gyr time scales. 
On the other hand, \citet{Ehlert2023MNRAS} perform simulations of self-regulated magnetized AGN feedback in cool-core clusters and see behavior that is more similar to our results -- i.e., the jets in their simulations, regardless of the precise details of jet momentum density or triggering mechanism, are continuously on.  An interesting intermediate case is provided by \citet{Yang2016ApJ}, who use cold-gas triggered hydrodynamical jets in an initially magnetized medium.  This work shows highly variable AGN energies but no indication of Gyr-scale patterns.  It is unclear whether the difference in jet behavior between hydrodynamical and magnetohydrodynamical simulations of self-regulated AGN feedback is primarily caused by the magnetic fields themselves or by other feedback parameters.  For example, the momentum density of the jet affects the jet behavior -- see, e.g., the discussion in Section 7.2 of \citet{Prasad2020ApJ}, Section 5.1 of \citet{Prasad2022ApJ}, and Section 3 of \citet{Weinberger2023MNRAS}.  In general, lighter jets are more easily contained at smaller cluster-centric radii and are also more easily deflected by cold gas than heavier jets. This may result in the simulated cluster's center displaying less variability in its thermodynamic state than for heavier jets.

Conversely, some studies have shown that jet precession or reorientation might play a role in the dynamics of the cold phase. In \citet{Beckmann_2019}, rapid reorientation of the jets on timescales of a few tens of Myr pushes the cold phase outwards in a nearly isotropic way, effectively shutting off the fueling of the AGN for a few tens of Myr and possibly resulting in the alternation between heating and cooling dominated phases.  

\subsection{On the cluster magnetic fields}
\label{sec:disc-bfields}

The calculations that include magnetic fields show that AGN feedback has a significant impact on the magnetic fields in our simulated clusters.  While the AGN injects a substantial amount of energy into the cluster only 1\% of it is magnetic energy, with the rest being distributed between kinetic (74\%) and thermal energy (25\%).
Over the duration of the simulation this means that for a jet with a sustained total energy injection rate of $10^{44}$ ($10^{45}$)~erg/s, the total magnetic energy injected into the system per billion years is $\simeq 3 \times 10^{58}$ ($3 \times 10^{59}$)~ergs.  The region of AGN influence is effectively within $\simeq 100$ kpc of the cluster center, and the total amount of magnetic energy contained within that volume is $\simeq 1.7 \times 10^{58}$\,ergs (\fid),  $\simeq 9.9 \times 10^{57}$\,ergs (\weakC), and  $\simeq 4.3 \times 10^{56}$\,ergs (\weakB) -- larger than the original magnetic energy contained within that region ($\simeq 4.9 \times 10^{57}$~ergs for the \fid/\weakC case; $\simeq 4.9 \times 10^{51}$~ergs for the \weakB case), but not larger than the amount of magnetic field energy injected by the AGN. 

The weak vs.~strong initial magnetic field simulations inject different amounts of magnetic energy -- roughly a factor of six integrated over the time range of $0.4-2.4$~Gyr.  Beyond this bulk difference in total energy injection,
the differences between the simulations shown in  Figure~\ref{fig:B-vs-r} are informative.
The ``strong B'' cases with different cooling rates are very similar in amplitude and shape, though the stronger cooling case has somewhat higher overall field strengths within 100 kpc at later times (by roughly a factor of two).
There is clear evidence of field amplification out to several tens of kpc.  This is an interesting contrast to the weak initial magnetic field case, where the fields are comparably large to the strong initial magnetic field case only within $\simeq 5$~kpc of the cluster center, and are much weaker at larger radii (and fall off tremendously outside of $\simeq 100$~kpc.

Since the amount of magnetic fields produced by the AGN in these three simulations are effectively the same, we
can draw two conclusions from this.  First, it seems likely that most of the magnetic field energy coming from the AGN jet is either deposited at large clustercentric radii or dissipated away.  The argument that much of the injected magnetic energy is deposited at large radii is supported by Figure~\ref{fig:theta-vs-r}, which shows where the ``tracer fluid'' injected with the AGN can be found.  This fluid is deposited in the same cells where the magnetic field is deposited, and thus is a good proxy for where the injected fields ultimately reside. However, the ICM at large radii is much less dense than at low radii, and the flux-frozen magnetic fields will adiabatically lower in amplitude, making it difficult to differentiate between the origins.  It is also possible that the magnetic energy in the jet is turned into kinetic energy, and ultimately dissipated.  It is likely that this is a sub-dominant process, however, because the initial configuration of our injected magnetic field is a loop and thus it is unclear how the injected magnetic fields would exert force on the ICM to transfer magnetic energy into kinetic energy. 

In addition, we expect that the magnetic field is additionally amplified by small scale dynamo action \citep{Brandenburg2005,Rincon2019}.
This is particularly visible in the \weakB case in the magnetic energy spectra in Fig.~\ref{fig:spec-avg} where the magnetic spectrum peaks at $\simeq 10$\,kpc scales.
Thus, the strongest coherent field exists on this scale, which is significantly larger
than the injection scale.
At the same time, the fields are amplified $>100\times$ within $\simeq 100$\,kpc in the \weakB case,
indicating significant overall magnetic field amplification (see Fig.~\ref{fig:B-vs-r}).
Similarly, when comparing the simulations with identical initial field strength (\fid and \weakC), the specific kinetic energy spectra, i.e., the raw velocity fluctuations perfectly agree
on average (see left panel of Fig.~\ref{fig:spec-avg}), whereas there is consistently
less power in magnetic fluctuation on scales $\lesssim 20$\,kpc in the \weakC case compared
to the \fid case.
Assuming that the velocity fluctuations reached a stationary state in both cases, the
additional power injected by the AGN in the \fid case has been converted to magnetic energy
dynamically, i.e., eventually through dynamo action.
A future energy transfer analysis following~\citet{Grete2017PhPl} will be able
to quantify the different processes.

Finally, we note that the fields outside of $\simeq 100$\,kpc from the cluster core can be seen to weaken over time, as can be seen most clearly in the left and center panels of Figure~\ref{fig:B-vs-r}.  This is unlikely to come from significant evolution of the intracluster medium at those scales, since the cooling times at large scale are $\gg 1$~Gyr and the ICM entropy profile is stable (see the middle and top panels of Figure~\ref{fig:Ktcool-r}, respectively).  We note that the initial conditions in our MHD simulations include an initial magnetic field that is perturbed on scales between $50-200$~kpc (see Section~\ref{sec:initialV_B}), as well as initial velocity perturbations on the same scale.
In the strong initial field simulations the resulting Alfv\'{e}n Mach number is less than one at large radii, which means that the magnetic fields can readily exert force on the plasma via magnetic tension in that region of the cluster.  Since the initial field configuration was not designed to be initially force-free (i.e., $(\nabla \times \mathbf{B}) \times \mathbf{B} = 0$), it is likely that the magnetic field is exerting force on the plasma and transferring its energy into fluid motion.  In the weak initial field run the gas velocity is highly super-Alfv\'{e}nic  at large radii at all times, which means that it is difficult for this transfer of energy to occur in that region of the cluster.  Conclusive evidence for this is phenomenon is challenging to obtain because fluid motion is observed in all of the simulations at large radii (see the top panel of Figure~\ref{fig:MsMapb-r}) at late times. This is difficult to disentangle from kinetic energy deposited by the AGN jet, and further analysis is required.

\subsection{On the impact of the AGN jet}
\label{sec:agn-jet-impact}

Figure~\ref{fig:theta-vs-r} shows the location of tracer fluid from the AGN as a function of radius and polar angle, both at a specific time and averaged over a range of time.  The instantaneous and time-averaged results are broadly consistent with each other and show where the AGN plasma has been mixed in with the ambient medium.  In the central region of the cluster ($r \lesssim 10$~kpc) we see that this tracer fluid is relatively isotropically distributed, but there is relatively little of it -- most of the jet material ends up at large cluster radii.  At those radii the tracer fluid in the stronger initial field cases is constrained to be in the polar regions of the cluster
(i.e., along the z-axis of the simulation, which is the jet launching axis), and out to $\simeq 300$~kpc in both cases.  The jet material ends up at quite large radii, consistent with it being launched with a high entropy and then buoyantly rising for some distance.

The \weakB and  \hydro simulations have similar  behavior at small radii, but the tracer fluid  does not reach as far from the centre as the \fid and \weakC cases.  The difference between the different initial field strengths are striking. In both cases, the jets are magnetized at identical fractional levels although the total amount of magnetic energy is scaled to the AGN power, which differs by a factor of $\simeq 6$. However, the behavior of the \weakB run is quite similar to the \hydro run.  This suggests that the relatively strong magnetic field in the ambient medium ($\sim 1~\mu$G in the strong field case compared to $0.01~\mu$G in the weak field case) suppresses mixing of the  jet material and thus facilitates its buoyant rise to large radii.  This is plausible, because even though the magnetic field is globally weak at large radii (with $\beta \simeq 10^3$ at large radii in the strong field case)  magnetic tension can still be a significant factor on small scales.
However, further analysis is required to disentangle the causalities given the differences
in AGN power over time.

\subsection{On the origin and evolution of cold gas}
\label{sec:disc-filaments}

While all of the simulations have cold gas in them, the simulations with stronger magnetic fields have cold gas that goes out to significantly larger radii.  Figure~\ref{fig:T-vs-r} shows that significant amounts of cold gas can be seen out to $\simeq 80$~kpc in the case with the strongest cooling and out to $\simeq 40$~kpc in the weaker cooling case. We emphasize that our simulations are run with homogeneous and constant metallicity and that this assumption in the strong cooling runs does not hold in reality, as metallicity in the Perseus cluster is observed to reach values below $\sim$ 0.5 Z$_\odot$ for radii larger than $\sim$ 50 kpc \citep{Sanders_2007}. Assuming Solar metallicity everywhere in the simulated box likely results in overestimated cooling at larger radii. In the weak initial field and hydrodynamical case (both with strong cooling) cold gas is largely constrained to be within a rotating disk with a radius of $\simeq 5-6$~kpc.   The bottom panel of Figure~\ref{fig:power_vs_time} suggests that the \textit{total amount} of cold gas is also impacted by the magnetic fields -- the high magnetic field run with strong cooling has $3-4\times$ more cold gas than the weak field run with the same level of cooling, and $8-10\times$ the cold gas seen in the hydrodyanmical run, from $T=1.5-2.5$~Gyr. This run also has a factor of a few more cold gas than the equivalent run with weaker cooling. From these results we infer that magnetic fields both promote the formation of cold gas and allow it to be much more widely distributed in the cluster than in cases with negligible or absent fields.  Morphologically, simulations with strong fields tend to develop cold ``filaments,'' possibly aligned with the local magnetic field structure. In \cite{Fournier2024A&A} this behavior is studied in some detail, and we will continue to investigate this topic in a following paper.

\subsection{Computational challenges and solutions}
\label{sec:disc-comp-approach}

While the \athenapk  code is highly performant, highly scalable, and based on a codebase that is extensively used for massively parallel simulations on GPU-based computers \citep[i.e., the Parthenon framework;][]{Parthenon}, moving to a completely new computing platform typically exposes new challenges.  This is particularly true when moving to a platform such as the Frontier exascale supercomputer \citep{Atchley2023} at the Oak Ridge Leadership Computing Facility (OLCF), which at the time we began to run our calculations on it in mid-2023 was the largest supercomputer in the world\footnote{see \url{https://top500.org/lists/top500/2023/11/}}.  Frontier, unlike most other large GPU-based supercomputers, uses the AMD ``Instinct'' MI250X GPUs rather than NVIDIA GPUs, has an HPE Slingshot interconnect, and has a site-wide 700 PB Lustre filesystem (``Orion'') that has its own large compute cluster to manage I/O and metadata for Frontier and OLCF's other supercomputers.  This means that, in addition to running \athenapk simulations at a scale more than an order of magnitude larger than on previous systems, we were doing so with very new (and thus relatively untested) computing, networking, and file storage hardware and their associated hardware drivers, compilers, MPI and parallel I/O libraries, and other related software.  This resulted in several specific challenges that had to be solved, which we detail below; in addition, we direct the interested reader to \citet{cug23} for a more detailed analysis of performance variability of \athenapk on Frontier.

The most notable challenge we ran into was related to the Lustre file system. When we began this project \athenapk used the HDF5 library \citep{HDF5-10.1145/1966895.1966900}, which uses MPI-IO for parallel reading and writing of files \citep{MPI-IO-10.1145/301816.301826}. While this gave perfectly acceptable results on smaller systems (such as Frontier's $\simeq 200$~petaflop predecessor, Summit) we found that HDF5+MPI-IO was both extremely slow on Frontier (taking more than half an hour to write an output file using the default filesystem and MPI-IO configuration) and very unstable
-- simulations would crash with very little feedback or silently stall, wasting allocation time\footnote{As a temporary workaround to silent stalls in IO wasting allocation time, we added logic to our job scripts to monitor activity in \texttt{STDOUT}. If nothing was written to \texttt{STDOUT} for a minute, we made the job script terminate itself.}. After extensive experimentation with environmental variable we found that disabling collective buffering, disabling the ``self extending layout'' and setting a fixed stripe size (to 2/3 of the maximum OSTs) resulted in acceptable (though not ideal) performance, i.e., writing an output file in $\mathcal{O}(3)$ minutes, in cases where the write was successful and not stalling.
One key issue of our IO approach remained though: writing to a single file.
To mitigate this (in the long run) we added support for
the OpenPMD file format \citep{osti_1880636}, which can make use of the ADIOS2 scalable I/O framework \citep{ADIOS2-10.1007/978-3-030-96498-6_6}.  ADIOS2 implements a streaming data pipeline methodology that effectively decouples the simulation code itself from the I/O subsystem, allowing much greater flexibility than HDF5+MPI-IO.  Furthermore, the ADIOS2 framework is well-matched to the design of modern supercomputer file systems.  As a consequence of this change we removed I/O performance as a bottleneck in future simulations.

In an attempt to reduce the volume of data output and the associated analysis burden, we added support for using the \textsc{Ascent} in-situ visualization library \citep{Larsen_2022} to \athenapk, but encountered multiple issues with performance and correctness. Since \ascent internally uses \textsc{HDF5}, using \ascent to produce slices of our simulations as they were running created an I/O bottleneck that we were unable to resolve. Additionally, support for ghost zones is inconsistent internally throughout \ascent, so we were unable to generate correct histograms or spherically-averaged profiles using this library. We also found that performance of several \ascent components were significantly worse on AMD GPUs as compared to NVIDIA GPUs, but we were unable to find a workaround for these issues.
This led to the addition of calculating histograms in-situ in \parthenon itself.

\subsection{Limitations}
\label{sec:disc-limitations}

As with all numerical simulations, the work presented in this paper (and in the following papers in this series) have limitations.  The most important limitations are discussed in the following subsections:

\subsubsection{Missing Energy Sources}

Aside from the AGN at the cluster center and the volumetric thermal feedback from the central galaxy's stellar population, there are no additional sources of energy injection in these simluations.  That means that we are missing several potential sources of energy that might stimulate bulk motion and turbulence, and ultimately heat the intracluster medium.  These missing sources originate from the hierarchical growth of cosmic structure. That is, we are missing the potential impact of recent cluster mergers (e.g., shocks and ``sloshing'' in the ICM), accretion of material (including galaxies and filamentary plasma) from the cosmic web, motions of non-central galaxies in the intracluster medium, and more generally, the motion of any massive substructure in the cluster.  While these dynamical energy sources significantly contribute to a galaxy cluster's overall energy budget over its lifetime, it is unlikely that any of them dominate the heating rate in a cluster's innermost regions (i.e., in the region where we focus our efforts).  For example, \citet{Fielding2020ApJ} compare both cosmological and idealized simulations of Milky Way-mass galaxies and find results that are similar at small fractions of the virial radii, but differ significantly at larger radii, indicating that dynamical heating less important than feedback heating at small radii. One would expect this trend to continue at larger halo mass scales, with deeper gravitational potential wells constraining the relevance of idealized simulations such as ours to even smaller fractions of the virial radius.

\subsubsection{Plasma Physics}

We neglect the impact of plasma physics beyond the assumptions built into the equations of ideal magnetohydrodynamics (MHD).  We do not include cosmic ray populations, which could potentially be a significant component of the ICM energy budget in certain subvolumes (such as AGN jet-driven bubbles) as well as valuable sources of observational information from, e.g., radio emission from cosmic ray electrons \citep[see, e.g.,][for a recent review]{2023A&ARv..31....4R}. We are also missing the impact of non-ideal MHD effects relating to the weak collisionality of the ICM plasma, such as anisotropic viscosity, conduction, and the local generation of magnetic fields.  The consequences of this physics have been extensively studied both analytically and numerically \citep[see, e.g.,][]{Schekochihin2009ApJS,Kunz2012ApJ,ZuHone2015ApJ,StOnge2018ApJ,Squire2023JPlPh,Zhou2024ApJ}.  Inclusion of additional plasma physics is a clear avenue for future work.  We are also explicitly ignoring both special and general relativity when modeling the impact of AGN jet feedback in group and cluster environments.  This is a reasonable approximation because the spatial and temporal scales that we are modeling are orders of magnitude larger than those relevant to, e.g., general relativistic MHD simulations of black hole accretion and jet launching \citep[see, e.g.,][]{Liska2020MNRAS,Kaaz2023ApJ}, and by the time the launched jets approach the scales that we are modeling they are weakly relativistic at most (i.e., $\frac{v}{c} \ll 1$).

\subsubsection{Stellar Feedback}

The algorithms that we use to approximate star formation and supernova mass and energy feedback in the central cluster galaxy (Sec.~\ref{sec:stellar-feedback}) are relatively simple, and in particular are volumetric rather than particle-based in nature.  In addition, feedback that approximates Type II supernovae originating from cold, dense gas in the halo center is approximated as instantaneous feedback rather than including a physically-appropriate delay.  We include feedback using this volumetric method because both theory and previous simulations have shown that it is potentially important for the self-regulation of the circumgalactic media in massive galaxies \citep[see, e.g.,][]{Voit2020ApJ,Prasad2020ApJ}. Still stellar feedback is globally sub-dominant.  It is unclear if having a more granular stellar feedback algorithm, which might impact the cold gas dynamics in the central region and, thus, the AGN fueling, would make a significant difference at the spatial and mass resolutions we are considering (in contrast to, e.g., at the dwarf or Milky Way-galaxy scale).  This is also an area of potential future study.

\subsection{Future papers in this series}
\label{sec:disc-future-work}

This paper is the first in a series of publications presenting the results from our simulation campaign, and presents a high-level overview of a subset of the simulations.  In the following papers we will perform detailed analyses of specific aspects of the simulations and examine a broader range of galaxy groups and clusters than is presented in this paper.
First results on the velocity structure functions (VSFs) of the multiphase ICM and the effects of projections (at various viewing angles and times) have already been published in \citet{Fournier2025A&A}.
Key findings include that there is no correlation between the hot and cold phase VSFs and that that observational biases such as projection effects, atmospheric seeing and the viewing angle cannot be ignored when interpreting line-of-sight VSF.
In addition, we plan to do the following:

We will examine the mechanisms by which the magnetized AGN jets inject energy and drive turbulence in the intracluster medium, and will examine the detailed properties of that turbulence using the energy transfer analysis tools that we have developed \citep{Grete2017PhPl}.  This extends previous work done by this collaboration on driven MHD turbulence \citep[e.g.,][]{Grete2021ApJ,Grete2023ApJ} into more physically-realistic environments.

We will create synthetic observations of the H$\alpha$ and X-ray-emitting plasma in our simulated clusters and compare this to galaxy cluster observations that are probing the dynamical state of local clusters \citep[e.g.,][]{Li2020ApJ,deVries2023MNRAS}.  In particular, we intend to examine the relationship between turbulent power spectra, line-of-sight velocity structure functions, X-ray surface brightness fluctuations, and potential systematic differences between the behavior seen at the resolution of the simulations themselves vs.~what can be discerned via measurements made with current observatories. 

We will use the full simulation suite, which includes galaxy groups with significantly different entropy profiles \citep[motivated by][]{Prasad2020ApJ,Prasad2022ApJ} and galaxy clusters spanning the mass range $10^{14} - 10^{15}$~M$_\odot$ to examine the self-regulating behavior of magnetized AGN in this range of environments using a uniform set of physics and numerical methods.

We will compare subvolumes extracted from our calculations with more idealized numerical experiments in order to understand systematic differences between them, and thus address challenges in connecting both to observations.  Our intent is to initially focus on comparisons with setups of hydrodynamic and magnetohydrodynamic turbulence in stratified media and the generation of multiphase structures
\citep[e.g.,][]{Ji2018MNRAS,Mohapatra2022MNRAS,Wibking2024arXiv241003886W}, but ultimately we will extend these comparisons to a broader range of phenomena.

It is also our intention to make this simulation data publicly available within a reasonable time period.  The primary challenge relates to data volume -- the raw data produced by our simulation campaign exceeds 4 petabytes, and it will take time to process this data to reduce its overall volume, to generate metadata, and to archive it in a place where it can be acquired.  We anticipate that this will require: (1) excising the scientifically interesting subvolumes of simulations from the full-scale calculations; (2) reducing the overall precision of the datasets produced (likely by using a lossy data compression method); (3) extensively documenting these steps as well as all simulation parameter files, codes, and output files; and (4) working with support staff at the relevant national computing centers to ensure that these datasets are available in a findable and searchable way.

\section{Conclusions and Outlook}\label{sec:conclusions}


This is the first paper in a series presenting results on the first exascale simulations examining the self-regulation of idealized galaxy clusters with magnetized jets.  The key results from this work are as follows:

\begin{enumerate}

  \item The idealized exascale MHD cluster simulations in this study self-regulate.  For the duration of the simulations the cluster core maintains a roughly constant thermodynamic state including significant amounts of cold gas (M$_{\rm cold} \sim 10^{10} M_\odot$).  The AGN injects energy at variable rates of $\simeq 10^{44}-10^{45}$~erg/s, with no discernible on/off duty cycle as is often seen in comparable hydrodynamic simulations. This is in agreement with radio observations of massive galaxy clusters \citep{2019A&A...622A..17S}.  Despite having no precession of the AGN jet we observe variability in the jets' alignment as well as episodic behavior due to the interaction of the jets with cold gas in the cluster center.

\item Cold gas filaments are not monolithic; rather, they have complex structures. The interaction of the jet with the cold gas leads to appreciable deflection of the jets, as discussed in detail in \cite{Fournier2024A&A}. Our simulations also show that AGN jets form large cavities in the ICM, some of which are off-axis (deflected away from jet axis) due to this interaction.

\item Magnetic fields significantly impact the formation and distribution of cold gas. 
Simulations initialized with stronger magnetic fields ($B=1~\mu$G) show enhanced cold gas formation out to much larger radii ($r\sim100$ kpc) compared to hydro simulations or those initialized with weak magnetic fields ($B=0.001~\mu$G) where cold gas is typically seen within $r<20$ kpc.
The total amount of cold gas that is produced depends on both the cooling function (i.e., the assumed ICM metallicity) and the initial magnetic field strength. 
It is comparable between hydrodynamic simulation, the one with weak magnetic fields ($B=0.001~\mu$G), and the simulation with strong magnetic fields ($B=1~\mu$G) but a lower-metallicity ($0.3$~Z$_\odot$) cooling function.

\item The AGN jet drives the amplification of magnetic fields in the cluster cores. Magnetic field amplification is locally fast (more than $100\times$ in less than 100\,Myr on the smallest scales) compared to global time scales (i.e., the sound-crossing time scale of the cluster core).

\item Both the kinetic and magnetic energy power spectra of turbulence in the cluster core grows over the first $\sim 1$~Gyr of AGN jet activity and ultimately achieves a roughly steady state. Moverover, the fluctuations in the power spectra amplitude and slope are linked to the AGN jet energy injection rate.

\item The AGN feedback `sphere of influence' is limited to the cluster core in all simulations  (i.e., it is contained within a cluster-centric radius of $\simeq 100$~kpc). 
Within the `sphere of influence' ($r\sim 100$ kpc) AGN drive turbulence that leads to magnetic field amplification and and the formation of filaments of cold gas. The entropy profiles of the ICM show the impact of the cooling and feedback cycle within the cluster out to $\simeq 100$~kpc. Simulations with lower  metallicity and weaker magnetic fields tend to have more radially-concentrated cold gas and magnetic fields. 

\item A passive tracer fluid injected with the AGN shows that the effect of the AGN on the ICM is confined to the polar regions of the cluster at relatively large radii ($\gtrsim 10$~kpc).  The jet-polluted material in the more highly magnetized simulations tend to stay in the polar regions and can be found at larger radii (out to $\simeq 300$~kpc) compared to simulations with weak or no magnetic fields, which have a more uniform (albeit still polar) and centrally-concentrated ($\lesssim 120$~kpc) distribution in their affected regions.  In contrast, the turbulence driven by the AGN is volume-filling within the cluster core, which demonstrates that the impact of the AGN can be felt outside of the region that is directly mixed with the jet ejecta.

\end{enumerate}

In the following papers in this series we will examine the mechanisms by which the magnetized AGN jets inject energy and drive turbulence in the intracluster medium, as well as show the detailed properties of that turbulence (including energy transfer). Moreover, we will
make detailed comparisons with multi-wavelength observations. Finally, we will 
examine the self-regulating behavior of magnetized AGN in galaxy group as well as in clusters of a range of masses
and compare subvolumes extracted from our calculations with more idealized setups of, e.g., turbulence in stratified media.

\begin{acknowledgments}

The authors thank Greg Bryan, Hui Li, Yuan Li, Ricarda Beckmann, and
Rainer Weinberger for insightful
discussions and the Oak Ridge Leadership Computing Facility
support staff, in particular John Holmen, for incredibly valuable support at all stages of the
  INCITE project that resulted in this and following publications. BDW thanks Michael Zingale for useful discussions of performance, scaling, and bugs on \textsc{Frontier}.
PG acknowledges the support by the DFG grant SCHW 1358/5-1 within
the priority program SPP 1992.
BWO acknowledges support from NSF grants \#1908109 and \#2106575, NASA ATP grants NNX15AP39G and 80NSSC18K1105, and NASA TCAN grant 80NSSC21K1053.  BWO also thanks the Los Alamos National Laboratory Center for Nonlinear Studies for support via the Ulam Sabbatical Fellowship.
This research used resources of the Oak Ridge Leadership Computing Facility at the Oak Ridge National Laboratory, which is supported by the Office of Science of the U.S. Department of Energy under Contract No. DE-AC05-00OR22725. These resources were provided by as part of the DOE INCITE Leadership Computing Program under allocation AST-146 (PI: Brian O'Shea).\\
MF and MB acknowledge funding by the Deutsche Forschungsgemeinschaft (DFG,
German Research Foundation) under Germany’s Excellence Strategy – EXC
2121 “Quantum Universe” – 390833306. This research was supported in part by grant NSF PHY-2309135 to the Kavli Institute for Theoretical Physics (KITP).
DP is supported by the Royal Society through UKRI grant RF-ERE-210263 (PI: Freeke van de Voort).
All simulations were performed using the public MHD code
{\href{https://github.com/parthenon-hpc-lab/athenapk}{\scshape{AthenaPK}}},
which makes use of the
\href{https://github.com/kokkos/kokkos}{{\scshape{Kokkos}}} library \citep{Trott2022}  and the \href{https://github.com/parthenon-hpc-lab/parthenon}{{\scshape{Parthenon}}} adaptive mesh refinement framework \citep{Parthenon}. All data analysis was performed with \href{https://yt-project.org/}{{\scshape{yt}}} \citep{Turk_2011,yt4}, \href{https://matplotlib.org/}{{\scshape{Matplotlib}}} \citep{Matplotlib}, and  \href{https://numpy.org/}{{\scshape{Numpy}}} \citep{Numpy}. We thank their authors for making these software and packages publicly available.
\end{acknowledgments}

\bibliography{paper}{}

\begin{thebibliography}{}
\expandafter\ifx\csname natexlab\endcsname\relax\def\natexlab#1{#1}\fi
\providecommand{\url}[1]{\href{#1}{#1}}
\providecommand{\dodoi}[1]{doi:~\href{http://doi.org/#1}{\nolinkurl{#1}}}
\providecommand{\doeprint}[1]{\href{http://ascl.net/#1}{\nolinkurl{http://ascl.net/#1}}}
\providecommand{\doarXiv}[1]{\href{https://arxiv.org/abs/#1}{\nolinkurl{https://arxiv.org/abs/#1}}}

\bibitem[{Atchley {et~al.}(2023)Atchley, Zimmer, Lange, Bernholdt,
  Melesse~Vergara, Beck, Brim, Budiardja, Chandrasekaran, Eisenbach, Evans,
  Ezell, Frontiere, Georgiadou, Glenski, Grete, Hamilton, Holmen, Huebl,
  Jacobson, Joubert, Mcmahon, Merzari, Moore, Myers, Nichols, Oral,
  Papatheodore, Perez, Rogers, Schneider, Vay, \& Yeung}]{Atchley2023}
Atchley, S., Zimmer, C., Lange, J., {et~al.} 2023, in Proceedings of the
  International Conference for High Performance Computing, Networking, Storage
  and Analysis, SC '23 (New York, NY, USA: Association for Computing
  Machinery), \dodoi{10.1145/3581784.3607089}

\bibitem[{{Beattie} {et~al.}(2024){Beattie}, {Federrath}, {Klessen}, {Cielo},
  \& {Bhattacharjee}}]{Beattie2024arXiv240516626B}
{Beattie}, J.~R., {Federrath}, C., {Klessen}, R.~S., {Cielo}, S., \&
  {Bhattacharjee}, A. 2024, arXiv e-prints, arXiv:2405.16626,
  \dodoi{10.48550/arXiv.2405.16626}

\bibitem[{Beckmann {et~al.}(2019)Beckmann, Dubois, Guillard, Salome, Olivares,
  Polles, Cadiou, Combes, Hamer, Lehnert, \& Pineau~des Forets}]{Beckmann_2019}
Beckmann, R.~S., Dubois, Y., Guillard, P., {et~al.} 2019, Astronomy \&amp;
  Astrophysics, 631, A60, \dodoi{10.1051/0004-6361/201936188}

\bibitem[{{Biava} {et~al.}(2021){Biava}, {Brienza}, {Bonafede}, {Gitti},
  {Bonnassieux}, {Harwood}, {Edge}, {Riseley}, \& {Vantyghem}}]{Biava2021A&A}
{Biava}, N., {Brienza}, M., {Bonafede}, A., {et~al.} 2021, \aap, 650, A170,
  \dodoi{10.1051/0004-6361/202040063}

\bibitem[{{B{\^\i}rzan} {et~al.}(2004){B{\^\i}rzan}, {Rafferty}, {McNamara},
  {Wise}, \& {Nulsen}}]{Birzan2004ApJ}
{B{\^\i}rzan}, L., {Rafferty}, D.~A., {McNamara}, B.~R., {Wise}, M.~W., \&
  {Nulsen}, P.~E.~J. 2004, \apj, 607, 800, \dodoi{10.1086/383519}

\bibitem[{Booth \& Schaye(2009)}]{Booth2009}
Booth, C.~M., \& Schaye, J. 2009, Monthly Notices of the Royal Astronomical
  Society, 398, 53, \dodoi{10.1111/j.1365-2966.2009.15043.x}

\bibitem[{Brandenburg \& Subramanian(2005)}]{Brandenburg2005}
Brandenburg, A., \& Subramanian, K. 2005, Physics Reports, 417, 1 ,
  \dodoi{10.1016/j.physrep.2005.06.005}

\bibitem[{{Cavagnolo} {et~al.}(2008){Cavagnolo}, {Donahue}, {Voit}, \&
  {Sun}}]{Cavagnolo2008ApJ}
{Cavagnolo}, K.~W., {Donahue}, M., {Voit}, G.~M., \& {Sun}, M. 2008, \apjl,
  683, L107, \dodoi{10.1086/591665}

\bibitem[{Cavagnolo {et~al.}(2009)Cavagnolo, Donahue, Voit, \&
  Sun}]{cavagnoloIntraclusterMediumEntropy2009}
Cavagnolo, K.~W., Donahue, M., Voit, G.~M., \& Sun, M. 2009, ApJS, 182, 12,
  \dodoi{10.1088/0067-0049/182/1/12}

\bibitem[{{Cavagnolo} {et~al.}(2010){Cavagnolo}, {McNamara}, {Nulsen},
  {Carilli}, {Jones}, \& {B{\^\i}rzan}}]{Cavagnolo2010ApJ}
{Cavagnolo}, K.~W., {McNamara}, B.~R., {Nulsen}, P.~E.~J., {et~al.} 2010, \apj,
  720, 1066, \dodoi{10.1088/0004-637X/720/2/1066}

\bibitem[{{Chadayammuri} {et~al.}(2021){Chadayammuri}, {Tremmel}, {Nagai},
  {Babul}, \& {Quinn}}]{Mila2021}
{Chadayammuri}, U., {Tremmel}, M., {Nagai}, D., {Babul}, A., \& {Quinn}, T.
  2021, \mnras, 504, 3922, \dodoi{10.1093/mnras/stab1010}

\bibitem[{{Churazov} {et~al.}(2001){Churazov}, {Br{\"u}ggen}, {Kaiser},
  {B{\"o}hringer}, \& {Forman}}]{Churazov2001}
{Churazov}, E., {Br{\"u}ggen}, M., {Kaiser}, C.~R., {B{\"o}hringer}, H., \&
  {Forman}, W. 2001, \apj, 554, 261, \dodoi{10.1086/321357}

\bibitem[{{de Vries} {et~al.}(2023){de Vries}, {Mantz}, {Allen}, {Morris},
  {Zhuravleva}, {Canning}, {Ehlert}, {Ogorza{\l}ek}, {Simionescu}, \&
  {Werner}}]{deVries2023MNRAS}
{de Vries}, M., {Mantz}, A.~B., {Allen}, S.~W., {et~al.} 2023, \mnras, 518,
  2954, \dodoi{10.1093/mnras/stac3285}

\bibitem[{Dedner {et~al.}(2002)Dedner, Kemm, Kröner, Munz, Schnitzer, \&
  Wesenberg}]{Dedner2002}
Dedner, A., Kemm, F., Kröner, D., {et~al.} 2002, Journal of Computational
  Physics, 175, 645 , \dodoi{10.1006/jcph.2001.6961}

\bibitem[{{Dekel} \& {Birnboim}(2008)}]{Dekel2008}
{Dekel}, A., \& {Birnboim}, Y. 2008, \mnras, 383, 119,
  \dodoi{10.1111/j.1365-2966.2007.12569.x}

\bibitem[{{Donahue} \& {Voit}(2022)}]{DonahueVoit2022PhR}
{Donahue}, M., \& {Voit}, G.~M. 2022, \physrep, 973, 1,
  \dodoi{10.1016/j.physrep.2022.04.005}

\bibitem[{{Edge}(2001)}]{Edge2001MNRAS}
{Edge}, A.~C. 2001, \mnras, 328, 762, \dodoi{10.1046/j.1365-8711.2001.04802.x}

\bibitem[{{Ehlert} {et~al.}(2023){Ehlert}, {Weinberger}, {Pfrommer}, {Pakmor},
  \& {Springel}}]{Ehlert2023MNRAS}
{Ehlert}, K., {Weinberger}, R., {Pfrommer}, C., {Pakmor}, R., \& {Springel}, V.
  2023, \mnras, 518, 4622, \dodoi{10.1093/mnras/stac2860}

\bibitem[{{Fabian}(1994)}]{Fabian_1994}
{Fabian}, A.~C. 1994, \araa, 32, 277,
  \dodoi{10.1146/annurev.aa.32.090194.001425}

\bibitem[{{Fabian} {et~al.}(1984){Fabian}, {Nulsen}, \&
  {Canizares}}]{Fabian1984Nature}
{Fabian}, A.~C., {Nulsen}, P.~E.~J., \& {Canizares}, C.~R. 1984, \nat, 310,
  733, \dodoi{10.1038/310733a0}

\bibitem[{{Federrath} {et~al.}(2021){Federrath}, {Klessen}, {Iapichino}, \&
  {Beattie}}]{Federrath2021NatAs}
{Federrath}, C., {Klessen}, R.~S., {Iapichino}, L., \& {Beattie}, J.~R. 2021,
  Nature Astronomy, 5, 365, \dodoi{10.1038/s41550-020-01282-z}

\bibitem[{{Field}(1965)}]{Field1965ApJ}
{Field}, G.~B. 1965, \apj, 142, 531, \dodoi{10.1086/148317}

\bibitem[{{Fielding} {et~al.}(2020){Fielding}, {Tonnesen}, {DeFelippis}, {Li},
  {Su}, {Bryan}, {Kim}, {Forbes}, {Somerville}, {Battaglia}, {Schneider}, {Li},
  {Choi}, {Hayward}, \& {Hernquist}}]{Fielding2020ApJ}
{Fielding}, D.~B., {Tonnesen}, S., {DeFelippis}, D., {et~al.} 2020, \apj, 903,
  32, \dodoi{10.3847/1538-4357/abbc6d}

\bibitem[{Folk {et~al.}(2011)Folk, Heber, Koziol, Pourmal, \&
  Robinson}]{HDF5-10.1145/1966895.1966900}
Folk, M., Heber, G., Koziol, Q., Pourmal, E., \& Robinson, D. 2011, in
  Proceedings of the EDBT/ICDT 2011 Workshop on Array Databases, AD '11 (New
  York, NY, USA: Association for Computing Machinery), 36–47,
  \dodoi{10.1145/1966895.1966900}

\bibitem[{{Fournier} {et~al.}(2024){Fournier}, {Grete}, {Br{\"u}ggen},
  {Glines}, \& {O'Shea}}]{Fournier2024A&A}
{Fournier}, M., {Grete}, P., {Br{\"u}ggen}, M., {Glines}, F.~W., \& {O'Shea},
  B.~W. 2024, \aap, 691, A239, \dodoi{10.1051/0004-6361/202451031}

\bibitem[{{Fournier} {et~al.}(2025){Fournier}, {Grete}, {Br{\"u}ggen},
  {O'Shea}, {Prasad}, {Wibking}, {Glines}, \& {Mohapatra}}]{Fournier2025A&A}
{Fournier}, M., {Grete}, P., {Br{\"u}ggen}, M., {et~al.} 2025, \aap,
  \dodoi{10.1051/0004-6361/202554278}

\bibitem[{Ganguly {et~al.}(2023)Ganguly, Li, Olivares, Su, Combes, Prakash,
  Hamer, Guillard, \& Ha}]{Ganguly2023}
Ganguly, S., Li, Y., Olivares, V., {et~al.} 2023, The Nature of the Motions of
  Multiphase Filaments in the Centers of Galaxy Clusters.
\newblock \doarXiv{2304.09879}

\bibitem[{{Gaspari} {et~al.}(2013){Gaspari}, {Ruszkowski}, \&
  {Oh}}]{Gaspari2013MNRAS}
{Gaspari}, M., {Ruszkowski}, M., \& {Oh}, S.~P. 2013, \mnras, 432, 3401,
  \dodoi{10.1093/mnras/stt692}

\bibitem[{{Gaspari} {et~al.}(2012){Gaspari}, {Ruszkowski}, \&
  {Sharma}}]{Gaspari2012ApJ}
{Gaspari}, M., {Ruszkowski}, M., \& {Sharma}, P. 2012, \apj, 746, 94,
  \dodoi{10.1088/0004-637X/746/1/94}

\bibitem[{{Gingras} {et~al.}(2024){Gingras}, {Coil}, {McNamara}, {Perrotta},
  {Brighenti}, {Russell}, {Li}, {Oh}, \& {Ning}}]{Gingras2024ApJ}
{Gingras}, M.-J., {Coil}, A.~L., {McNamara}, B.~R., {et~al.} 2024, \apj, 977,
  159, \dodoi{10.3847/1538-4357/ad822a}

\bibitem[{{Grete} {et~al.}(2021){Grete}, {O'Shea}, \&
  {Beckwith}}]{Grete2021ApJ}
{Grete}, P., {O'Shea}, B.~W., \& {Beckwith}, K. 2021, \apj, 909, 148,
  \dodoi{10.3847/1538-4357/abdd22}

\bibitem[{{Grete} {et~al.}(2023){Grete}, {O'Shea}, \&
  {Beckwith}}]{Grete2023ApJ}
---. 2023, \apjl, 942, L34, \dodoi{10.3847/2041-8213/acaea7}

\bibitem[{{Grete} {et~al.}(2017){Grete}, {O'Shea}, {Beckwith}, {Schmidt}, \&
  {Christlieb}}]{Grete2017PhPl}
{Grete}, P., {O'Shea}, B.~W., {Beckwith}, K., {Schmidt}, W., \& {Christlieb},
  A. 2017, Physics of Plasmas, 24, 092311, \dodoi{10.1063/1.4990613}

\bibitem[{Grete {et~al.}(2023)Grete, Dolence, Miller, Brown, Ryan, Gaspar,
  Glines, Swaminarayan, Lippuner, Solomon, Shipman, Junghans, Holladay, Stone,
  \& Roberts}]{Parthenon}
Grete, P., Dolence, J.~C., Miller, J.~M., {et~al.} 2023, The International
  Journal of High Performance Computing Applications, 37, 465,
  \dodoi{10.1177/10943420221143775}

\bibitem[{Guo {et~al.}(2024)Guo, Stone, Quataert, \& Kim}]{Guo_2024}
Guo, M., Stone, J.~M., Quataert, E., \& Kim, C.-G. 2024, The Astrophysical
  Journal, 973, 141, \dodoi{10.3847/1538-4357/ad5fe7}

\bibitem[{{Hamer} {et~al.}(2014){Hamer}, {Edge}, {Swinbank}, {Oonk}, {Mittal},
  {McNamara}, {Russell}, {Bremer}, {Combes}, {Fabian}, {Nesvadba}, {O'Dea},
  {Baum}, {Salom{\'e}}, {Tremblay}, {Donahue}, {Ferland}, \&
  {Sarazin}}]{Hamer2014}
{Hamer}, S.~L., {Edge}, A.~C., {Swinbank}, A.~M., {et~al.} 2014, \mnras, 437,
  862, \dodoi{10.1093/mnras/stt1949}

\bibitem[{Harris {et~al.}(2020)Harris, Millman, van~der Walt, Gommers,
  Virtanen, Cournapeau, Wieser, Taylor, Berg, Smith, Kern, Picus, Hoyer, van
  Kerkwijk, Brett, Haldane, del R{\'{i}}o, Wiebe, Peterson,
  G{\'{e}}rard-Marchant, Sheppard, Reddy, Weckesser, Abbasi, Gohlke, \&
  Oliphant}]{Numpy}
Harris, C.~R., Millman, K.~J., van~der Walt, S.~J., {et~al.} 2020, Nature, 585,
  357, \dodoi{10.1038/s41586-020-2649-2}

\bibitem[{Holmen {et~al.}(2024)Holmen, Grete, \& Melesse~Vergara}]{cug23}
Holmen, J.~K., Grete, P., \& Melesse~Vergara, V.~G. 2024, Concurrency and
  Computation: Practice and Experience, 36, e8069,
  \dodoi{https://doi.org/10.1002/cpe.8069}

\bibitem[{{Hudson} {et~al.}(2010){Hudson}, {Mittal}, {Reiprich}, {Nulsen},
  {Andernach}, \& {Sarazin}}]{Hudson_2010}
{Hudson}, D.~S., {Mittal}, R., {Reiprich}, T.~H., {et~al.} 2010, \aap, 513,
  A37, \dodoi{10.1051/0004-6361/200912377}

\bibitem[{Huebl {et~al.}(2015)Huebl, Lehe, Kuschel, Fortmann-Grote, Koller, \&
  USDOE}]{osti_1880636}
Huebl, A., Lehe, R., Kuschel, S., {et~al.} 2015, The openPMD Standard,
  \dodoi{10.11578/dc.20220810.25}

\bibitem[{{Hunter}(2007)}]{Matplotlib}
{Hunter}, J.~D. 2007, Computing in Science and Engineering, 9, 90,
  \dodoi{10.1109/MCSE.2007.55}

\bibitem[{{Ji} {et~al.}(2018){Ji}, {Oh}, \& {McCourt}}]{Ji2018MNRAS}
{Ji}, S., {Oh}, S.~P., \& {McCourt}, M. 2018, \mnras, 476, 852,
  \dodoi{10.1093/mnras/sty293}

\bibitem[{{Kaaz} {et~al.}(2023){Kaaz}, {Murguia-Berthier}, {Chatterjee},
  {Liska}, \& {}}]{Kaaz2023ApJ}
{Kaaz}, N., {Murguia-Berthier}, A., {Chatterjee}, K., {Liska}, M. T.~P., \& {},
  A. 2023, \apj, 950, 31, \dodoi{10.3847/1538-4357/acc7a1}

\bibitem[{{Kunz} {et~al.}(2012){Kunz}, {Bogdanovi{\'c}}, {Reynolds}, \&
  {Stone}}]{Kunz2012ApJ}
{Kunz}, M.~W., {Bogdanovi{\'c}}, T., {Reynolds}, C.~S., \& {Stone}, J.~M. 2012,
  \apj, 754, 122, \dodoi{10.1088/0004-637X/754/2/122}

\bibitem[{Larsen {et~al.}(2022)Larsen, Brugger, Childs, \&
  Harrison}]{Larsen_2022}
Larsen, M., Brugger, E., Childs, H., \& Harrison, C. 2022, in {In Situ
  Visualization For Computational Science} (Cham, Switzerland: Mathematics and
  Visualization book series from Springer Publishing), 255 -- 279

\bibitem[{Li \& Bryan(2014)}]{Li_2014}
Li, Y., \& Bryan, G.~L. 2014, The Astrophysical Journal, 789, 54,
  \dodoi{10.1088/0004-637x/789/1/54}

\bibitem[{{Li} {et~al.}(2015){Li}, {Bryan}, {Ruszkowski}, {Voit}, {O'Shea}, \&
  {Donahue}}]{Li2015ApJ}
{Li}, Y., {Bryan}, G.~L., {Ruszkowski}, M., {et~al.} 2015, \apj, 811, 73,
  \dodoi{10.1088/0004-637X/811/2/73}

\bibitem[{{Li} {et~al.}(2020){Li}, {Gendron-Marsolais}, {Zhuravleva}, {Xu},
  {Simionescu}, {Tremblay}, {Lochhaas}, {Bryan}, {Quataert}, {Murray},
  {Boselli}, {Hlavacek-Larrondo}, {Zheng}, {Fossati}, {Li}, {Emsellem},
  {Sarzi}, {Arzamasskiy}, \& {Vishniac}}]{Li2020ApJ}
{Li}, Y., {Gendron-Marsolais}, M.-L., {Zhuravleva}, I., {et~al.} 2020, \apjl,
  889, L1, \dodoi{10.3847/2041-8213/ab65c7}

\bibitem[{{Liska} {et~al.}(2020){Liska}, {Tchekhovskoy}, \&
  {Quataert}}]{Liska2020MNRAS}
{Liska}, M., {Tchekhovskoy}, A., \& {Quataert}, E. 2020, \mnras, 494, 3656,
  \dodoi{10.1093/mnras/staa955}

\bibitem[{Mathews {et~al.}(2006)Mathews, Faltenbacher, \&
  Brighenti}]{Mathews2006}
Mathews, W.~G., Faltenbacher, A., \& Brighenti, F. 2006, \apj, 638, 659,
  \dodoi{10.1086/499119}

\bibitem[{{McCourt} {et~al.}(2012){McCourt}, {Sharma}, {Quataert}, \&
  {Parrish}}]{McCourt2012MNRAS}
{McCourt}, M., {Sharma}, P., {Quataert}, E., \& {Parrish}, I.~J. 2012, \mnras,
  419, 3319, \dodoi{10.1111/j.1365-2966.2011.19972.x}

\bibitem[{{McDonald} {et~al.}(2019){McDonald}, {Allen}, {Hlavacek-Larrondo},
  {Mantz}, {Bayliss}, {Benson}, {Brodwin}, {Bulbul}, {Canning}, {Chiu},
  {Forman}, {Garmire}, {Gupta}, {Khullar}, {Mohr}, {Reichardt}, \&
  {Schrabback}}]{McDonald_2019}
{McDonald}, M., {Allen}, S.~W., {Hlavacek-Larrondo}, J., {et~al.} 2019, \apj,
  870, 85, \dodoi{10.3847/1538-4357/aaf394}

\bibitem[{{McNamara} {et~al.}(2014){McNamara}, {Russell}, {Nulsen}, {Edge},
  {Murray}, {Main}, {Vantyghem}, {Combes}, {Fabian}, {Salome}, {Kirkpatrick},
  {Baum}, {Bregman}, {Donahue}, {Egami}, {Hamer}, {O'Dea}, {Oonk}, {Tremblay},
  \& {Voit}}]{McNamara2014ApJ}
{McNamara}, B.~R., {Russell}, H.~R., {Nulsen}, P.~E.~J., {et~al.} 2014, \apj,
  785, 44, \dodoi{10.1088/0004-637X/785/1/44}

\bibitem[{{Meece} {et~al.}(2017){Meece}, {Voit}, \& {O'Shea}}]{Meece2017ApJ}
{Meece}, G.~R., {Voit}, G.~M., \& {O'Shea}, B.~W. 2017, \apj, 841, 133,
  \dodoi{10.3847/1538-4357/aa6fb1}

\bibitem[{Miyoshi \& Kusano(2005)}]{Miyoshi2005}
Miyoshi, T., \& Kusano, K. 2005, Journal of Computational Physics, 208, 315 ,
  \dodoi{10.1016/j.jcp.2005.02.017}

\bibitem[{{Mohapatra} {et~al.}(2022){Mohapatra}, {Federrath}, \&
  {Sharma}}]{Mohapatra2022MNRAS}
{Mohapatra}, R., {Federrath}, C., \& {Sharma}, P. 2022, \mnras, 514, 3139,
  \dodoi{10.1093/mnras/stac1610}

\bibitem[{{Mohapatra} {et~al.}(2023){Mohapatra}, {Sharma}, {Federrath}, \&
  {Quataert}}]{Mohapatra2023MNRAS}
{Mohapatra}, R., {Sharma}, P., {Federrath}, C., \& {Quataert}, E. 2023, \mnras,
  525, 3831, \dodoi{10.1093/mnras/stad2574}

\bibitem[{{Naab} \& {Ostriker}(2017)}]{Naab2017ARA&A}
{Naab}, T., \& {Ostriker}, J.~P. 2017, \araa, 55, 59,
  \dodoi{10.1146/annurev-astro-081913-040019}

\bibitem[{Navarro {et~al.}(1997)Navarro, Frenk, \&
  White}]{navarroUniversalDensityProfile1997}
Navarro, J.~F., Frenk, C.~S., \& White, S. D.~M. 1997, ApJ, 490, 493,
  \dodoi{10.1086/304888}

\bibitem[{{Nelson} {et~al.}(2024){Nelson}, {Pillepich}, {Ayromlou}, {Lee},
  {Lehle}, {Rohr}, \& {Truong}}]{Nelson2024A&A}
{Nelson}, D., {Pillepich}, A., {Ayromlou}, M., {et~al.} 2024, \aap, 686, A157,
  \dodoi{10.1051/0004-6361/202348608}

\bibitem[{{Nobels} {et~al.}(2022){Nobels}, {Schaye}, {Schaller}, {Bah{\'e}}, \&
  {Chaikin}}]{Nobels2022MNRAS}
{Nobels}, F. S.~J., {Schaye}, J., {Schaller}, M., {Bah{\'e}}, Y.~M., \&
  {Chaikin}, E. 2022, \mnras, 515, 4838, \dodoi{10.1093/mnras/stac2061}

\bibitem[{{O'Dea} {et~al.}(2008){O'Dea}, {Baum}, {Privon}, {Noel-Storr},
  {Quillen}, {Zufelt}, {Park}, {Edge}, {Russell}, {Fabian}, {Donahue},
  {Sarazin}, {McNamara}, {Bregman}, \& {Egami}}]{ODea_2008}
{O'Dea}, C.~P., {Baum}, S.~A., {Privon}, G., {et~al.} 2008, \apj, 681, 1035,
  \dodoi{10.1086/588212}

\bibitem[{{Olivares} {et~al.}(2019){Olivares}, {Salome}, {Combes}, {Hamer},
  {Guillard}, {Lehnert}, {Polles}, {Beckmann}, {Dubois}, {Donahue}, {Edge},
  {Fabian}, {McNamara}, {Rose}, {Russell}, {Tremblay}, {Vantyghem}, {Canning},
  {Ferland}, {Godard}, {Peirani}, \& {Pineau des Forets}}]{Olivares2019A&A}
{Olivares}, V., {Salome}, P., {Combes}, F., {et~al.} 2019, \aap, 631, A22,
  \dodoi{10.1051/0004-6361/201935350}

\bibitem[{{Oosterloo} {et~al.}(2024){Oosterloo}, {Morganti}, \&
  {Murthy}}]{Oosterloo_2024}
{Oosterloo}, T., {Morganti}, R., \& {Murthy}, S. 2024, Nature Astronomy, 8,
  256, \dodoi{10.1038/s41550-023-02138-y}

\bibitem[{{O'Shea} {et~al.}(2005){O'Shea}, {Nagamine}, {Springel}, {Hernquist},
  \& {Norman}}]{OShea2005ApJS}
{O'Shea}, B.~W., {Nagamine}, K., {Springel}, V., {Hernquist}, L., \& {Norman},
  M.~L. 2005, \apjs, 160, 1, \dodoi{10.1086/432645}

\bibitem[{{Panagoulia} {et~al.}(2014){Panagoulia}, {Fabian}, {Sanders}, \&
  {Hlavacek-Larrondo}}]{Panagoulia2014MNRAS}
{Panagoulia}, E.~K., {Fabian}, A.~C., {Sanders}, J.~S., \& {Hlavacek-Larrondo},
  J. 2014, \mnras, 444, 1236, \dodoi{10.1093/mnras/stu1499}

\bibitem[{{Pellissier} {et~al.}(2023){Pellissier}, {Hahn}, \&
  {Ferrari}}]{Pellissier2023MNRAS}
{Pellissier}, A., {Hahn}, O., \& {Ferrari}, C. 2023, \mnras, 522, 721,
  \dodoi{10.1093/mnras/stad888}

\bibitem[{{Peterson} {et~al.}(2003){Peterson}, {Kahn}, {Paerels}, {Kaastra},
  {Tamura}, {Bleeker}, {Ferrigno}, \& {Jernigan}}]{Peterson2003ApJ}
{Peterson}, J.~R., {Kahn}, S.~M., {Paerels}, F.~B.~S., {et~al.} 2003, \apj,
  590, 207, \dodoi{10.1086/374830}

\bibitem[{{Piotrowska} {et~al.}(2022){Piotrowska}, {Bluck}, {Maiolino}, \&
  {Peng}}]{2022MNRAS.512.1052P}
{Piotrowska}, J.~M., {Bluck}, A. F.~L., {Maiolino}, R., \& {Peng}, Y. 2022,
  \mnras, 512, 1052, \dodoi{10.1093/mnras/stab3673}

\bibitem[{{Pizzolato} \& {Soker}(2005)}]{Pizzolato2005ApJ}
{Pizzolato}, F., \& {Soker}, N. 2005, \apj, 632, 821, \dodoi{10.1086/444344}

\bibitem[{Poeschel {et~al.}(2022)Poeschel, E, Godoy, Podhorszki, Klasky,
  Eisenhauer, Davis, Wan, Gainaru, Gu, Koller, Widera, Bussmann, \&
  Huebl}]{ADIOS2-10.1007/978-3-030-96498-6_6}
Poeschel, F., E, J., Godoy, W.~F., {et~al.} 2022, in Driving Scientific and
  Engineering Discoveries Through the Integration of Experiment, Big Data, and
  Modeling and Simulation, ed. J.~Nichols, A.~B. Maccabe, J.~Nutaro,
  S.~Pophale, P.~Devineni, T.~Ahearn, \& B.~Verastegui (Cham: Springer
  International Publishing), 99--118

\bibitem[{{Prasad} {et~al.}(2015){Prasad}, {Sharma}, \&
  {Babul}}]{Prasad2015ApJ}
{Prasad}, D., {Sharma}, P., \& {Babul}, A. 2015, ApJ, 811, 108,
  \dodoi{10.1088/0004-637X/811/2/108}

\bibitem[{{Prasad} {et~al.}(2022){Prasad}, {Voit}, \& {O'Shea}}]{Prasad2022ApJ}
{Prasad}, D., {Voit}, G.~M., \& {O'Shea}, B.~W. 2022, \apj, 932, 18,
  \dodoi{10.3847/1538-4357/ac69ee}

\bibitem[{{Prasad} {et~al.}(2020){Prasad}, {Voit}, {O'Shea}, \&
  {Glines}}]{Prasad2020ApJ}
{Prasad}, D., {Voit}, G.~M., {O'Shea}, B.~W., \& {Glines}, F. 2020, \apj, 905,
  50, \dodoi{10.3847/1538-4357/abc33c}

\bibitem[{Prunier {et~al.}(2024)Prunier, Hlavacek-Larrondo, Pillepich, Lehle,
  \& Nelson}]{Prunier_2024}
Prunier, M., Hlavacek-Larrondo, J., Pillepich, A., Lehle, K., \& Nelson, D.
  2024, X-ray cavities in TNG-Cluster: AGN phenomena in the full cosmological
  context.
\newblock \doarXiv{2410.21366}

\bibitem[{Qiu {et~al.}(2019)Qiu, Bogdanović, Li, Park, \& Wise}]{Qiu_2019}
Qiu, Y., Bogdanović, T., Li, Y., Park, K., \& Wise, J.~H. 2019, The
  Astrophysical Journal, 877, 47, \dodoi{10.3847/1538-4357/ab18fd}

\bibitem[{{Rafferty} {et~al.}(2008){Rafferty}, {McNamara}, \&
  {Nulsen}}]{Rafferty2008ApJ}
{Rafferty}, D.~A., {McNamara}, B.~R., \& {Nulsen}, P.~E.~J. 2008, \apj, 687,
  899, \dodoi{10.1086/591240}

\bibitem[{{Rafferty} {et~al.}(2006){Rafferty}, {McNamara}, {Nulsen}, \&
  {Wise}}]{Rafferty2006ApJ}
{Rafferty}, D.~A., {McNamara}, B.~R., {Nulsen}, P.~E.~J., \& {Wise}, M.~W.
  2006, \apj, 652, 216, \dodoi{10.1086/507672}

\bibitem[{Riffel {et~al.}(2020)Riffel, Storchi-Bergmann, Zakamska, \&
  Riffel}]{Riffel2020}
Riffel, R.~A., Storchi-Bergmann, T., Zakamska, N.~L., \& Riffel, R. 2020,
  \mnras, 496, 4857, \dodoi{10.1093/mnras/staa1922}

\bibitem[{Rincon(2019)}]{Rincon2019}
Rincon, F. 2019, Journal of Plasma Physics, 85, 205850401,
  \dodoi{10.1017/S0022377819000539}

\bibitem[{{Russell} {et~al.}(2014){Russell}, {McNamara}, {Edge}, {Nulsen},
  {Main}, {Vantyghem}, {Combes}, {Fabian}, {Murray}, {Salom{\'e}}, {Wilman},
  {Baum}, {Donahue}, {O'Dea}, {Oonk}, {Tremblay}, \& {Voit}}]{Russell2014ApJ}
{Russell}, H.~R., {McNamara}, B.~R., {Edge}, A.~C., {et~al.} 2014, \apj, 784,
  78, \dodoi{10.1088/0004-637X/784/1/78}

\bibitem[{{Ruszkowski} {et~al.}(2004){Ruszkowski}, {Br{\"u}ggen}, \&
  {Begelman}}]{2004ApJ...611..158R}
{Ruszkowski}, M., {Br{\"u}ggen}, M., \& {Begelman}, M.~C. 2004, \apj, 611, 158,
  \dodoi{10.1086/422158}

\bibitem[{{Ruszkowski} \& {Oh}(2011)}]{Ruszkowski2011}
{Ruszkowski}, M., \& {Oh}, S.~P. 2011, \mnras, 414, 1493,
  \dodoi{10.1111/j.1365-2966.2011.18482.x}

\bibitem[{{Ruszkowski} \& {Pfrommer}(2023)}]{2023A&ARv..31....4R}
{Ruszkowski}, M., \& {Pfrommer}, C. 2023, \aapr, 31, 4,
  \dodoi{10.1007/s00159-023-00149-2}

\bibitem[{{Sabater} {et~al.}(2019){Sabater}, {Best}, {Hardcastle}, {Shimwell},
  {Tasse}, {Williams}, {Br{\"u}ggen}, {Cochrane}, {Croston}, {de Gasperin},
  {Duncan}, {G{\"u}rkan}, {Mechev}, {Morabito}, {Prandoni}, {R{\"o}ttgering},
  {Smith}, {Harwood}, {Mingo}, {Mooney}, \& {Saxena}}]{2019A&A...622A..17S}
{Sabater}, J., {Best}, P.~N., {Hardcastle}, M.~J., {et~al.} 2019, \aap, 622,
  A17, \dodoi{10.1051/0004-6361/201833883}

\bibitem[{{Salom{\'e}} {et~al.}(2006){Salom{\'e}}, {Combes}, {Edge},
  {Crawford}, {Erlund}, {Fabian}, {Hatch}, {Johnstone}, {Sanders}, \&
  {Wilman}}]{Salome2006A&A}
{Salom{\'e}}, P., {Combes}, F., {Edge}, A.~C., {et~al.} 2006, \aap, 454, 437,
  \dodoi{10.1051/0004-6361:20054745}

\bibitem[{Sanders \& Fabian(2007)}]{Sanders_2007}
Sanders, J.~S., \& Fabian, A.~C. 2007, Monthly Notices of the Royal
  Astronomical Society, 381, 1381–1399,
  \dodoi{10.1111/j.1365-2966.2007.12347.x}

\bibitem[{{Schaye} {et~al.}(2023){Schaye}, {Kugel}, {Schaller}, {Helly},
  {Braspenning}, {Elbers}, {McCarthy}, {van Daalen}, {Vandenbroucke}, {Frenk},
  {Kwan}, {Salcido}, {Bah{\'e}}, {Borrow}, {Chaikin}, {Hahn}, {Hu{\v{s}}ko},
  {Jenkins}, {Lacey}, \& {Nobels}}]{Schaye2023MNRAS}
{Schaye}, J., {Kugel}, R., {Schaller}, M., {et~al.} 2023, \mnras, 526, 4978,
  \dodoi{10.1093/mnras/stad2419}

\bibitem[{{Schekochihin} {et~al.}(2009){Schekochihin}, {Cowley}, {Dorland},
  {Hammett}, {Howes}, {Quataert}, \& {Tatsuno}}]{Schekochihin2009ApJS}
{Schekochihin}, A.~A., {Cowley}, S.~C., {Dorland}, W., {et~al.} 2009, \apjs,
  182, 310, \dodoi{10.1088/0067-0049/182/1/310}

\bibitem[{Schure {et~al.}(2009)Schure, Kosenko, Kaastra, Keppens, \&
  Vink}]{schureNewRadiativeCooling2009a}
Schure, K.~M., Kosenko, D., Kaastra, J.~S., Keppens, R., \& Vink, J. 2009,
  Astronomy \& Astrophysics, 508, 751, \dodoi{10.1051/0004-6361/200912495}

\bibitem[{{Sharma} {et~al.}(2012){Sharma}, {McCourt}, {Quataert}, \&
  {Parrish}}]{Sharma2012}
{Sharma}, P., {McCourt}, M., {Quataert}, E., \& {Parrish}, I.~J. 2012, \mnras,
  420, 3174, \dodoi{10.1111/j.1365-2966.2011.20246.x}

\bibitem[{{Shimwell} {et~al.}(2022){Shimwell}, {Hardcastle}, {Tasse}, {Best},
  {R{\"o}ttgering}, {Williams}, {Botteon}, {Drabent}, {Mechev}, {Shulevski},
  {van Weeren}, {Bester}, {Br{\"u}ggen}, {Brunetti}, {Callingham}, {Chy{\.z}y},
  {Conway}, {Dijkema}, {Duncan}, {de Gasperin}, {Hale}, {Haverkorn}, {Hugo},
  {Jackson}, {Mevius}, {Miley}, {Morabito}, {Morganti}, {Offringa}, {Oonk},
  {Rafferty}, {Sabater}, {Smith}, {Schwarz}, {Smirnov}, {O'Sullivan},
  {Vedantham}, {White}, {Albert}, {Alegre}, {Asabere}, {Bacon}, {Bonafede},
  {Bonnassieux}, {Brienza}, {Bilicki}, {Bonato}, {Calistro Rivera}, {Cassano},
  {Cochrane}, {Croston}, {Cuciti}, {Dallacasa}, {Danezi}, {Dettmar}, {Di
  Gennaro}, {Edler}, {En{\ss}lin}, {Emig}, {Franzen}, {Garc{\'\i}a-Vergara},
  {Grange}, {G{\"u}rkan}, {Hajduk}, {Heald}, {Heesen}, {Hoang}, {Hoeft},
  {Horellou}, {Iacobelli}, {Jamrozy}, {Jeli{\'c}}, {Kondapally}, {Kukreti},
  {Kunert-Bajraszewska}, {Magliocchetti}, {Mahatma}, {Ma{\l}ek}, {Mandal},
  {Massaro}, {Meyer-Zhao}, {Mingo}, {Mostert}, {Nair}, {Nakoneczny},
  {Nikiel-Wroczy{\'n}ski}, {Orr{\'u}}, {Pajdosz-{\'S}mierciak}, {Pasini},
  {Prandoni}, {van Piggelen}, {Rajpurohit}, {Retana-Montenegro}, {Riseley},
  {Rowlinson}, {Saxena}, {Schrijvers}, {Sweijen}, {Siewert}, {Timmerman},
  {Vaccari}, {Vink}, {West}, {Wo{\l}owska}, {Zhang}, \&
  {Zheng}}]{Shimwell2022A&A}
{Shimwell}, T.~W., {Hardcastle}, M.~J., {Tasse}, C., {et~al.} 2022, \aap, 659,
  A1, \dodoi{10.1051/0004-6361/202142484}

\bibitem[{Simionescu {et~al.}(2011)Simionescu, Allen, Mantz, Werner, Takei,
  Morris, Fabian, Sanders, Nulsen, George, \& Taylor}]{Simionescu2011}
Simionescu, A., Allen, S.~W., Mantz, A., {et~al.} 2011, Science, 331, 1576,
  \dodoi{10.1126/science.1200331}

\bibitem[{{Soker} {et~al.}(2001){Soker}, {White}, {David}, \&
  {McNamara}}]{soker2001}
{Soker}, N., {White}, III, R.~E., {David}, L.~P., \& {McNamara}, B.~R. 2001,
  \apj, 549, 832, \dodoi{10.1086/319433}

\bibitem[{{Squire} {et~al.}(2023){Squire}, {Kunz}, {Arzamasskiy}, {Johnston},
  {Quataert}, \& {Schekochihin}}]{Squire2023JPlPh}
{Squire}, J., {Kunz}, M.~W., {Arzamasskiy}, L., {et~al.} 2023, Journal of
  Plasma Physics, 89, 905890417, \dodoi{10.1017/S0022377823000727}

\bibitem[{{St-Onge} \& {Kunz}(2018)}]{StOnge2018ApJ}
{St-Onge}, D.~A., \& {Kunz}, M.~W. 2018, \apjl, 863, L25,
  \dodoi{10.3847/2041-8213/aad638}

\bibitem[{Stone {et~al.}(2020)Stone, Tomida, White, \&
  Felker}]{stoneAthenaAdaptiveMesh2020}
Stone, J.~M., Tomida, K., White, C.~J., \& Felker, K.~G. 2020, The
  Astrophysical Journal Supplement Series, 249, 4,
  \dodoi{10.3847/1538-4365/ab929b}

\bibitem[{{Sun} {et~al.}(2009){Sun}, {Voit}, {Donahue}, {Jones}, {Forman}, \&
  {Vikhlinin}}]{Sun2009ApJ}
{Sun}, M., {Voit}, G.~M., {Donahue}, M., {et~al.} 2009, \apj, 693, 1142,
  \dodoi{10.1088/0004-637X/693/2/1142}

\bibitem[{Thakur {et~al.}(1999)Thakur, Gropp, \&
  Lusk}]{MPI-IO-10.1145/301816.301826}
Thakur, R., Gropp, W., \& Lusk, E. 1999, in Proceedings of the Sixth Workshop
  on I/O in Parallel and Distributed Systems, IOPADS '99 (New York, NY, USA:
  Association for Computing Machinery), 23–32, \dodoi{10.1145/301816.301826}

\bibitem[{Townsend(2009)}]{Townsend2009}
Townsend, R. H.~D. 2009, The Astrophysical Journal Supplement Series, 181, 391,
  \dodoi{10.1088/0067-0049/181/2/391}

\bibitem[{{Tremblay} {et~al.}(2018){Tremblay}, {Combes}, {Oonk}, {Russell},
  {McDonald}, {Gaspari}, {Husemann}, {Nulsen}, {McNamara}, {Hamer}, {O'Dea},
  {Baum}, {Davis}, {Donahue}, {Voit}, {Edge}, {Blanton}, {Bremer}, {Bulbul},
  {Clarke}, {David}, {Edwards}, {Eggerman}, {Fabian}, {Forman}, {Jones},
  {Kerman}, {Kraft}, {Li}, {Powell}, {Randall}, {Salom{\'e}}, {Simionescu},
  {Su}, {Sun}, {Urry}, {Vantyghem}, {Wilkes}, \& {ZuHone}}]{Tremblay2018ApJ}
{Tremblay}, G.~R., {Combes}, F., {Oonk}, J.~B.~R., {et~al.} 2018, \apj, 865,
  13, \dodoi{10.3847/1538-4357/aad6dd}

\bibitem[{Trott {et~al.}(2022)Trott, Lebrun-Grandié, Arndt, Ciesko, Dang,
  Ellingwood, Gayatri, Harvey, Hollman, Ibanez, Liber, Madsen, Miles,
  Poliakoff, Powell, Rajamanickam, Simberg, Sunderland, Turcksin, \&
  Wilke}]{Trott2022}
Trott, C.~R., Lebrun-Grandié, D., Arndt, D., {et~al.} 2022, IEEE Transactions
  on Parallel and Distributed Systems, 33, 805,
  \dodoi{10.1109/TPDS.2021.3097283}

\bibitem[{{Turk} {et~al.}(2011){Turk}, {Smith}, {Oishi}, {Skory}, {Skillman},
  {Abel}, \& {Norman}}]{Turk_2011}
{Turk}, M.~J., {Smith}, B.~D., {Oishi}, J.~S., {et~al.} 2011, The Astrophysical
  Journal Supplement Series, 192, 9, \dodoi{10.1088/0067-0049/192/1/9}

\bibitem[{{Turk} {et~al.}(2024){Turk}, {Smith}, {Oishi}, {Skory}, {Skillman},
  {Abel}, \& {Norman}}]{yt4}
---. 2024, {Introducing yt 4.0: Analysis and Visualization of Volumetric Data},
  \url{https://yt-project.github.io/yt-4.0-paper/}

\bibitem[{{Vantyghem} {et~al.}(2014){Vantyghem}, {McNamara}, {Russell}, {Main},
  {Nulsen}, {Wise}, {Hoekstra}, \& {Gitti}}]{Vantyghem2014MNRAS}
{Vantyghem}, A.~N., {McNamara}, B.~R., {Russell}, H.~R., {et~al.} 2014, \mnras,
  442, 3192, \dodoi{10.1093/mnras/stu1030}

\bibitem[{{Vantyghem} {et~al.}(2021){Vantyghem}, {McNamara}, {O'Dea}, {Baum},
  {Combes}, {Edge}, {Fabian}, {McDonald}, {Nulsen}, {Russell}, \&
  {Salom{\'e}}}]{Vantyghem2021ApJ}
{Vantyghem}, A.~N., {McNamara}, B.~R., {O'Dea}, C.~P., {et~al.} 2021, \apj,
  910, 53, \dodoi{10.3847/1538-4357/abe306}

\bibitem[{{Voigt} \& {Fabian}(2004)}]{Voigt2004}
{Voigt}, L.~M., \& {Fabian}, A.~C. 2004, \mnras, 347, 1130,
  \dodoi{10.1111/j.1365-2966.2004.07285.x}

\bibitem[{{Voit} \& {Bryan}(2001)}]{voit2001Natur}
{Voit}, G.~M., \& {Bryan}, G.~L. 2001, \nat, 414, 425, \dodoi{10.1038/35106523}

\bibitem[{{Voit} {et~al.}(2008){Voit}, {Cavagnolo}, {Donahue}, {Rafferty},
  {McNamara}, \& {Nulsen}}]{Voit_2008}
{Voit}, G.~M., {Cavagnolo}, K.~W., {Donahue}, M., {et~al.} 2008, \apjl, 681,
  L5, \dodoi{10.1086/590344}

\bibitem[{Voit {et~al.}(2015)Voit, Donahue, O’Shea, Bryan, Sun, \&
  Werner}]{Voit2015a}
Voit, G.~M., Donahue, M., O’Shea, B.~W., {et~al.} 2015, The Astrophysical
  Journal Letters, 803, L21, \dodoi{10.1088/2041-8205/803/2/L21}

\bibitem[{{Voit} {et~al.}(2020){Voit}, {Bryan}, {Prasad}, {Frisbie}, {Li},
  {Donahue}, {O'Shea}, {Sun}, \& {Werner}}]{Voit2020ApJ}
{Voit}, G.~M., {Bryan}, G.~L., {Prasad}, D., {et~al.} 2020, \apj, 899, 70,
  \dodoi{10.3847/1538-4357/aba42e}

\bibitem[{Wang {et~al.}(2021)Wang, Ruszkowski, Pfrommer, Oh, \&
  Yang}]{Wang_2021}
Wang, C., Ruszkowski, M., Pfrommer, C., Oh, S.~P., \& Yang, H.-Y.~K. 2021,
  Monthly Notices of the Royal Astronomical Society, 504, 898–909,
  \dodoi{10.1093/mnras/stab966}

\bibitem[{Weinberger {et~al.}(2017)Weinberger, Ehlert, Pfrommer, Pakmor, \&
  Springel}]{Weinberger2017}
Weinberger, R., Ehlert, K., Pfrommer, C., Pakmor, R., \& Springel, V. 2017,
  Monthly Notices of the Royal Astronomical Society, 470, 4530,
  \dodoi{10.1093/mnras/stx1409}

\bibitem[{{Weinberger} {et~al.}(2018){Weinberger}, {Springel}, {Pakmor},
  {Nelson}, {Genel}, {Pillepich}, {Vogelsberger}, {Marinacci}, {Naiman},
  {Torrey}, \& {Hernquist}}]{2018MNRAS.479.4056W}
{Weinberger}, R., {Springel}, V., {Pakmor}, R., {et~al.} 2018, \mnras, 479,
  4056, \dodoi{10.1093/mnras/sty1733}

\bibitem[{{Weinberger} {et~al.}(2023){Weinberger}, {Su}, {Ehlert}, {Pfrommer},
  {Hernquist}, {Bryan}, {Springel}, {Li}, {Burkhart}, {Choi}, \&
  {Faucher-Gigu{\`e}re}}]{Weinberger2023MNRAS}
{Weinberger}, R., {Su}, K.-Y., {Ehlert}, K., {et~al.} 2023, \mnras, 523, 1104,
  \dodoi{10.1093/mnras/stad1396}

\bibitem[{{Wibking} {et~al.}(2024){Wibking}, {Voit}, \&
  {O'Shea}}]{Wibking2024arXiv241003886W}
{Wibking}, B.~D., {Voit}, G.~M., \& {O'Shea}, B.~W. 2024, arXiv e-prints,
  arXiv:2410.03886, \dodoi{10.48550/arXiv.2410.03886}

\bibitem[{{Yang} \& {Reynolds}(2016)}]{Yang2016ApJ}
{Yang}, H. Y.~K., \& {Reynolds}, C.~S. 2016, \apj, 818, 181,
  \dodoi{10.3847/0004-637X/818/2/181}

\bibitem[{{Zhou} {et~al.}(2024){Zhou}, {Zhdankin}, {Kunz}, {Loureiro}, \&
  {Uzdensky}}]{Zhou2024ApJ}
{Zhou}, M., {Zhdankin}, V., {Kunz}, M.~W., {Loureiro}, N.~F., \& {Uzdensky},
  D.~A. 2024, \apj, 960, 12, \dodoi{10.3847/1538-4357/ad0b0f}

\bibitem[{{ZuHone} {et~al.}(2015){ZuHone}, {Kunz}, {Markevitch}, {Stone}, \&
  {Biffi}}]{ZuHone2015ApJ}
{ZuHone}, J.~A., {Kunz}, M.~W., {Markevitch}, M., {Stone}, J.~M., \& {Biffi},
  V. 2015, \apj, 798, 90, \dodoi{10.1088/0004-637X/798/2/90}

\end{thebibliography}
\bibliographystyle{aasjournal}



\end{document}